# Prioritizing network communities

Marinka Zitnik,[1] Rok Sosič,[1] and Jure Leskovec[1,2,*]

[1] Computer Science Department, Stanford University, Stanford, CA, USA
[2] Chan Zuckerberg Biohub, San Francisco, CA, USA

[*] To whom correspondence should be addressed; E-mail: jure@cs.stanford.edu

**Uncovering modular structure in networks is fundamental for systems in biology, physics, and engineering. Community detection identifies candidate modules as hypotheses, which then need to be validated through experiments, such as mutagenesis in a biological laboratory. Only a few communities can typically be validated, and it is thus important to prioritize which communities to select for downstream experimentation. Here we develop** CRANK**, a mathematically principled approach for prioritizing network communities.** CRANK **efficiently evaluates robustness and magnitude of structural features of each community and then combines these features into the community prioritization.** CRANK **can be used with any community detection method. It needs only information provided by the network structure and does not require any additional metadata or labels. However, when available,** CRANK **can incorporate domain-specific information to further boost performance. Experiments on many large networks show that** CRANK **effectively prioritizes communities, yielding a nearly 50-fold improvement in community prioritization.**

Networks exhibit modular structure[1] and uncovering it is fundamental for advancing the understanding of complex systems across sciences[2,3]. Methods for community detection[4], also called node clustering or graph partitioning, allow for computational detection of modular structure by identifying a division of network's nodes into groups, also called communities[5–10]. Such communities provide predictions/hypotheses about potential modules of the network, which then need to be experimentally validated and confirmed. However, in large networks, community detection methods typically identify many thousands of communities[6,7] and only a small fraction can be rigorously tested and validated by follow-up experiments. For example, gene communities detected in a gene interaction network[11] provide predictions/hypotheses about disease pathways[2,3], but to confirm these predictions scientists have to test every detected community by performing experiments in a wet laboratory[3,8]. Because experimental validation of detected communities is resource-intensive and generally only a small number of communities can be investigated, one must prioritize the communities in order to choose which ones to investigate experimentally.

In the context of biological networks, several methods for community or cluster analysis have been developed[2,3,12–15]. However, these methods crucially rely and depend on knowledge in external databases, such as Gene Ontology (GO) annotations[16], protein domain databases, gene expression data, patient clinical profiles, and sequence information, in order to calculate the quality of communities derived from networks. Furthermore, they require this information to be available for all communities. This means that if genes in a given community are not present in a gene knowledge database then it is not possible for existing methods to even consider that community. This issue is exacerbated because knowledge databases are incomplete and biased toward better-studied genes[11]. Furthermore, these methods do not apply in domains at the frontier of science where domain-specific knowledge is scarce or non-existent, such as in the case of cell-cell similarity networks[17], microbiome networks[18,19], and chemical interaction networks[20]. Thus, there is a need for a general solution to prioritize communities based on network information only.

Here, we present CRank, a general approach that takes a network and detected communities as its input and produces a ranked list of communities, where high-ranking communities represent promising candidates for downstream experiments. CRank can be applied in conjunction with any community detection method (Supplementary Notes 2 and 5) and needs only the network structure, requiring no domain-specific meta or label information about the network. However, when domain-specific supervised information is available, CRANK can integrate this extra information to boost performance (Supplementary Notes 9 and 10). CRANK can thus prioritize communities that are well characterized in knowledge bases, such as GO annotations, as well as poorly characterized communities with limited or no annotations. Furthermore, CRANK is based on rigorous statistical methods to provide an overall rank for each detected community.

# Results

## Overview of CRANK community prioritization approach

CRANK community prioritization approach consists of the following steps (Figure 1). First, CRANK finds communities using an existing, preferred community detection method (Figure 1a). It then computes for each community four CRANK defined community prioritization metrics, which capture key structural features of the community (Figure 1b), and then it combines the community metrics via a aggregation method into a single overall score for each community (Figure 1c). Finally, CRANK prioritizes communities by ranking them by their decreasing overall score (Figure 1d).

CRANK uses four different metrics to characterize network connectivity features for each detected community (see Methods). These metrics evaluate the magnitude of structural features as well as their robustness against noise in the network structure. The rationale here is that high priority communities have high values of metrics and are also stable with respect to network perturbations. If a small change in the network structure—an edge added here, another deleted there—significantly changes the value of a prioritization metric then the community will not be considered high priority. We derive analytical expressions for calculating these metrics, which make CRANK computationally efficient and applicable to large networks (Supplementary Note 2). Because individual metrics may have different importance in different networks, a key element of CRANK is a rank aggregation method. This method combines the values of the four metrics into a single score for each community, which then determines the community's rank (see Methods and Supplementary Note 4). CRANK's aggregation method adjusts the impact of each metric on the ranking in a principled manner across different networks and also across different communities within a network, leading to robust rankings and a high-quality prioritization of communities (Supplementary Note 5).

### Synthetic networks

We first demonstrate CRANK by applying it to synthetic networks with planted community structure (Figure 2a). The goal of community prioritization is to identify communities that are most promising candidates for follow-up investigations. Since communities provide predictions about the modular structure of the network, promising candidates are communities that best correspond to the underlying modules. Thus, in this synthetic example, the aim of community prioritization can be seen as to rank communities based on how well they represent the underlying planted communities, while only utilizing information about network structure and without any additional information about the planted communities. We quantify prioritization quality by measuring the agreement between a ranked list of communities produced by CRANK and the gold standard ranking. In the gold standard ranking, communities are ordered in the decreasing order of how accurately each community reconstructs its corresponding planted community.

We experiment with random synthetic networks with planted community structure (Figure 2a), where we use a generic community detection method[7] to identify communities and then prioritize them using CRANK. We observe that CRANK produces correct prioritization—using only the unlabeled network structure, CRANK places communities that better correspond to planted communities towards the top of the ranking (Figure 2b), which indicates that CRANK can identify accurately detected communities by using the network structure alone and having no other data about planted community structure. Comparing the performance of CRANK to alternative ranking techniques, such as modularity[5] and conductance[21], we observe that CRANK performs 149% and 37% better than modularity and conductance, respectively, in terms of the Spearman's rank correlation between the generated ranking and the gold standard community ranking (Figure 2c). Moreover, we observed no correlation with the gold standard ranking when randomly ordering the detected communities. Although zero correlation is expected, poor performance of random ordering is especially illuminating because prioritization of communities is typically ignored in current network community studies.

### Networks of medical drugs with shared target proteins

Community rankings obtained by CRANK provide a rich source of testable hypotheses. For example, we consider a network of medical drugs where two drugs are connected if they share at least one target protein (Figure 3a). Because drugs that are used to treat closely related diseases tend to share target proteins[22], we expect that drugs belonging to the same community in the network will be rich in chemicals with similar therapeutic effects. Identification of these drug communities hence provides an attractive opportunity for finding new uses of drugs as well as for studying drugs' adverse effects[22].

After detecting drug communities using a standard community detection method[7], CRANK relies only on the network structure to prioritize the communities. We evaluate ranking performance by comparing it to metadata captured in external chemical databases and not used by the ranking method. We find that CRANK assigns higher priority to communities whose drugs are pharmacogenomically more similar (Figure 3b), indicating that higher-ranked communities contain drugs with more abundant drug-drug interactions, more similar chemical structure, and stronger textual associations. In contrast, ranking communities by modularity score gives a poor correspondence with information in the external chemical databases (Figure 3c).

We observe that the top ranked communities are composed from an unusual set of drugs (Figure 3a and Supplementary Table), yet drugs with unforeseen community assignment may represent novel candidates for drug repurposing[22]. Examining the highest ranked communities, we do not expect mifepristone, an abortifacient used in the first months of pregnancy, to appear together with a group of drugs used to treat inflammatory diseases. Another drug with unanticipated community assignment is minaprine, a psychotropic drug that is effective in the treatment of various depressive states[23]. Minaprine is an antidepressant that antagonizes behavioral despair; however, it shares target proteins with several cholinesterase inhibitors. Two examples of such inhibitors are physostigmine, used to treat glaucoma, and galantamine, a drug investigated for the treatment of moderate Alzheimer's disease[24]. In the case of minaprine, an antidepressant, it was just recently shown that this drug is also a cognitive enhancer that may halt the progression of Alzheimer's disease[25]. It is thus attractive that CRANK identified minaprine as a member of a community of primarily cholinesterase inhibitors, which suggests minaprine's potential for drug repurposing for Alzheimer's disease.

The analysis here was restricted to drugs approved for medical use by the U.S. Food and Drug Administration, because these drugs are accompanied by rich metadata that was used for evaluating community prioritization. We find that when CRANK integrates drug metadata into its prioritization model, CRANK can generate up to 55% better community rankings, even when the amount of additional information about drugs is small (Supplementary Note 10). However, approved medical drugs represent less than one percent of all small molecules with recorded interactions. Many of the remaining 99% of these molecules might be candidates for medical usage or drug repurposing but currently have little or no metadata in the chemical databases. This fact further emphasizes the need for methods such as CRANK that can prioritize communities based on network structure alone while not relying on any metadata in external chemical databases.

### Gene and protein interaction networks

CRANK can also prioritize communities in molecular biology networks, covering a spectrum of physical, genetic, and regulatory gene interactions[11]. In such networks, community detection is widely used because gene communities tend to correlate with cellular functions and thus provide hypotheses about biological pathways and protein complexes[2,3].

CRANK takes a network and communities detected in that network, and produces a rank-ordered list of communities. As before, while CRANK ranks the communities purely based on network structure, the external metadata about molecular functions, cellular components, and biological processes is used to assess the quality of the community ranking.

Considering highest ranked gene communities, CRANK's ranking contains on average 5 times more communities whose genes are significantly enriched for cellular functions, components, and processes[16] than random prioritization, and 13% more significantly enriched communities than modularity- or conductance-based ranking (Supplementary Note 11). For example, in the human protein-protein interaction network, the highest ranked community by CRANK is composed of 20 genes, including *PORCN*, *AQP5*, *FZD6*, *WNT1*, *WNT2*, *WNT3*, and other members of the Wnt signaling protein family[26] (Supplementary Note 11). Genes in that community form a biologically meaningful group that is functionally enriched in the Wnt signaling pathway processes (p-value = $6.4 \times 10^{-23}$), neuron differentiation (p-value = $1.6 \times 10^{-15}$), cellular response to retinoic acid (p-value = $2.9 \times 10^{-14}$), and in developmental processes (p-value = $9.2 \times 10^{-10}$).

Functional annotation of molecular networks is largely unavailable and incomplete, especially when studied objects are not genes but rather other entities, *e.g.*, miRNAs, mutations, single nucleotide variants, or genomic regions outside protein-coding loci[27]. Thus it is often not possible to simply rank the communities by their functional enrichment scores. In such scenarios, CRANK can prioritize communities reliably and accurately using only network structure without necessitat-ing any external databases. Gene communities that rank at the top according to CRANK represent predictions that could guide scientists to prioritize resource-intensive laboratory experiments.

**Megascale cell-cell similarity networks**

Single-cell RNA sequencing has transformed our understanding of complex cell populations[28]. While many types of questions can be answered using single-cell RNA-sequencing, a central focus is the ability to survey the diversity of cell types and composition of tissues within a sample of cells.

To demonstrate that CRANK scales to large networks, we used the single-cell RNA-seq dataset containing 1,306,127 embryonic mouse brain cells[29] for which no cell types are known. The dataset was preprocessed using standard procedures to select and filter the cells based on quality-control metrics, normalize and scale the data, detect highly variable genes, and remove unwanted sources of variation[9]. The dataset was represented as a weighted graph of nearest neighbor relations (edges) among cells (nodes), where relations indicated cells with similar gene expression patterns calculated using diffusion pseudotime analysis[30]. To partition this graph into highly interconnected communities we apply a community detection method proposed for single-cell data[8]. The method separates the cells into 141 fine-grained communities, the largest containing 18,788 (1.8% of) and the smallest only 203 (0.02% of) cells. After detecting the communities, CRANK takes the cell-cell similarity network and the detected communities, and generates a rank-ordered list of communities, assigning a priority to each community. CRANK's prioritization of communities derived from the cell-cell similarity network takes less than 2 minutes on a personal computer.

In the cell-cell similarity network, one could assume that top-ranked communities contain highly distinct marker genes[31], while low-ranked communities contain marker genes whose expression levels are spread out beyond cells in the community. To test this hypothesis, we identify marker genes for each detected community. In particular, for each community we find genes that are differentially expressed in the cells within the community[9] relative to all cells that are not in the community.

We find that high-ranked communities in CRANK contain cells with distinct marker genes, confirming the above hypothesis (average z-score of marker genes with respect to the bulk mean gene expression was above 200 and never smaller than 150) (Figure 4a-b). In contrast, cells in low-ranked communities show a weak expression activity diffused across the entire network and no community-specific expression activity (Figure 4c-d). Examining cells assigned to the highest-ranked community (rank 1 community) in CRANK, we find that most differentially expressed genes are *TYROBP*, *C1QB*, *C1QC*, *FCER1G*, and *C1QA* (at least a 200-fold difference in normalized expression with respect to the bulk mean expression[9]). It is known that these are immunoregulatory genes and that they play important roles in signal transduction in dendritic cells, osteoclasts, macrophages, and microglia[32]. In contrast, low-ranked communities (Figure 4 visualizes rank 139, rank 140, and rank 141 communities) contain predominantly cells in which genes show no community-specific expression. Genes in communities ranked lower by CRANK hence do not have localized mRNA expression levels, suggesting there are no good marker genes that define those communities[28]. Since the expression levels of mRNA are linked to cellular function and can be used to define cell types[28], the analysis here points to the potential of using highest-ranked communities in CRANK as candidates to characterize cells at the molecular level, even in datasets where no cells are yet classified into cell types.

**Analysis of CRANK prioritization approach**

The CRANK approach can be applied with any community detection method and can operate on directed, undirected, and weighted networks. Furthermore, CRANK can also use external domain-specific information to further boost prioritization performance (Supplementary Note 10). Results on diverse biological, information, and technological networks and on different community detection methods show that the second best performing approach changes considerably across networks, while CRANK always produces the best result, suggesting that it can effectively harness the network structure for community prioritization (Supplementary Note 8). CRANK automatically adjusts weights of the community metrics in the prioritization, resulting in each metric participating with different intensity across different networks (Supplementary Figure 6). This is in sharp contrast with deterministic approaches, which are negatively impacted by heterogeneity of network structures and network community models employed by different community detection methods. The four CRANK community prioritization metrics are essential and complementary. CRANK metrics considered together perform on average 45% better than the best single CRANK metric, and 26% better than any subset of three CRANK metrics (Supplementary Note 8). CRANK performs on average 38% better than approaches that combine alternative community metrics (Supplementary Note 8). Furthermore, CRANK can easily integrate any number of additional and domain-specific community metrics[2,12–15], and performs well in the presence of low-signal and noisy metrics (Supplementary Note 9). Furthermore, CRANK outperforms alternative approaches that combine the metrics by approximating NP-hard rank aggregation objectives (Supplementary Note 8).

## Discussion

The task of community prioritization is to rank-order communities detected by a community detection method such that communities with best prospects in downstream analysis are ranked towards the top. We demonstrated that prioritizing communities in biological, information, and technological networks is important for maximizing the yield of downstream analyses and experiments. Prior efforts crucially depend on external meta information to calculate the quality of communities with an additional constraint that this information has to be available for all communities. We devised a principled approach for the task of community prioritization. Although the approach does not need any meta information, it can utilize such information if it is available. Furthermore, CRANK is applicable even when the meta information is noisy, incomplete, or available only for a subset of communities.

The CRANK community ranking is based on the premise that high priority communities produce high values of community prioritization metrics and that these metrics are stable with respect to small perturbations of the network structure. Our findings support this premise and suggest that both the magnitude of the metrics and the robustness of underlying structural features have an important role in the performance of CRANK across a wide range of networks (Supplementary Note 8). CRANK can easily be extended using existing network metrics and can also consider new domain-specific scoring metrics (Supplementary Notes 9 and 10). Thus, it would be especially interesting to apply it to networks, where rich meta information exists and interesting domain-specific scoring metrics can be developed, such as protein interaction networks with disease pathway meta information[33], and molecular networks with genome-wide associations[34]. We believe that the CRANK approach opens the door to principled methods for prioritizing communities in large networks and, when coupled with experimental validation, can help us to speed-up scientific discovery process.

# Methods

## Community prioritization model

CRANK prioritizes communities based on the robustness and magnitude of multiple structural features of each community. For each feature $f$, we specify a corresponding prioritization metric $r_f$, which captures the magnitude and the robustness of $f$. Robustness of $f$ is defined as the change in the value of $f$ between the original network and its randomly perturbed version. The intent here is that high quality communities will have high values of $f$ and will also be robust to perturbations of the network structure. We define and discuss specific prioritization metrics later. Here, we first present the overall prioritization model.

Random perturbations of the network are based on rewiring of $\alpha$ fraction of the edges in a degree preserving manner[35] (Supplementary Note 2). Parameter $\alpha$ measures perturbation intensity; a value close to zero indicates that the network has only a few edges rewired whereas a value close to one corresponds to a maximally perturbed network, which is a random graph with the same degree distribution as the original network.

Even though the prioritization model is framed conceptually in terms of perturbing the network by rewiring its edges, CRANK never actually rewires the network when calculating the prioritization metrics. Network rewiring is a computationally expensive operation. Instead, we derive analytical expressions that evaluate the metrics in a closed form without physically perturbing the network (Supplementary Note 2), which leads to a substantial increase in scalability of CRANK.

Given structural feature $f$, we define prioritization metric $r_f$ to quantify the change in the value of $f$ between the original and the perturbed network. We want $r_f$ to capture the magnitude of feature $f$ in the original network as well as the change in the value of $f$ between the network and its perturbed version.

We define prioritization metric $r_f$ for community $C$ as:

$$r_f(C;\alpha) = \frac{f(C)}{1 + d_f(C,\alpha)}, \tag{1}$$

where $f(C)$ is the feature value of community $C$ in the original network, $\alpha$ measures perturbation intensity, $d_f(C,\alpha) = |f(C) - f(C|\alpha)|$ is the change of the feature value for community $C$ between the network and its $\alpha$-perturbed version, and $f(C|\alpha)$ is the value of feature $f$ in the $\alpha$-perturbed version of the network.

Generally, higher priority communities will have higher values of $r_f$. In particular, as $f$ can take values between zero and one, then $r_f$ also takes values between zero and one. $r_f$ attains value of zero for community $C$ whose value of $f(C)$ is zero. When $f(C)$ is nonzero, then $r_f(C;\alpha)$ down-weights it according to the sensitivity of community $C$ to network rewiring. $f(C)$ is down-weighted by the largest amount when it changes as much as possible under the network perturbation (*i.e.*, $d_f(C,\alpha) = 1$). And, $f(C)$ remains unchanged when community $C$ is maximally robust to network perturbation (*i.e.*, $d_f(C,\alpha) = 0$).

## Community prioritization metrics

Prioritization metric $r_f(C)$ captures the magnitude as well as the robustness of structural feature $f$ of community $C$. We define four different community prioritization metrics $r_f$. Through empirical analysis we show that these metrics holistically and non-redundantly quantify different features of network community structure (Supplementary Note 8). Each metric is necessary and contributes positively to the performance of CRANK. We combine these metrics into a global ranking of communities using a rank aggregation method that we describe later.

Given a network $G(\mathcal{V},\mathcal{E},\mathcal{C})$ with nodes $\mathcal{V}$, edges $\mathcal{E}$, and detected communities $\mathcal{C}$, CRANK can be applied in conjunction with any statistical community detection method that allows for computing the following three quantities: (1) the probability of node $u$ belonging to a given community $C$, $p_C(u) = p(u \in C)$, (2) the probability of an edge $p(u,v) = p((u,v) \in \mathcal{E})$, and (3) a contribution of community $C$ towards the existence of an edge $(u,v)$, $p_C(u,v) = p((u,v) \in \mathcal{E} | u,v \in C)$. Many commonly used community detection methods allow for computing the above three quantities (Supplementary Note 5).

Our rationale in defining the prioritization metrics is to measure properties that determine a high quality community, which is also robust and stable with respect to small perturbations of the network. For example, a genuine high quality community should provide good support for the existence of edges between its members in the original network as well as in the perturbed version. If a small change in the network structure—an edge added here, another deleted there—can completely change the value of the prioritization metric then the community should not be considered high quality. Analogously, a high quality community should have low confidence for edges pointing outside of the community both in the original as well as in the perturbed network.

**Community likelihood**

The community likelihood metric quantifies the overall connectivity of a given community. It measures the likelihood of the network structure induced by the nodes in the community. Note that the metric does not simply count the edges but considers them in a probabilistic way. As such it quantifies how well the observed edges can be explained by the community $C$. The intuition is that high quality community will contribute a large amount of likelihood to explain the observed edges.

We formalize the community likelihood for a given community $C$ as follows:

$$f_l(C|\alpha) = \prod_{u \in C} p_C(u) \prod_{v \in C} s_C(u, v|\alpha), \tag{2}$$

where $s_C(u, v|\alpha)$ is defined as follows:

$$s_C(u, v|\alpha) = \begin{cases} p_C(v)p_C(u, v|\alpha) & \text{if } (u, v) \in \mathcal{E} \\ p_C(v)(1 - p_C(u, v|\alpha)) & \text{if } (u, v) \notin \mathcal{E}. \end{cases}$$

Here, $p_C(u, v|\alpha)$ is a contribution of community $C$ towards the creation of edge $(u, v)$ under network perturbation intensity $\alpha$. We derive analytical expressions for $p_C(u, v|\alpha)$ which allows us to compute their values without ever actually perturbing the network (Supplementary Note 2).

Here (and for the other three prioritization metrics) we evaluate the feature in the original network ($f_l(C) = f_l(C|\alpha = 0)$) as well as in the slightly perturbed version of the network ($f_l(C|\alpha = 0.15)$). We then combine the two scores using the prioritization metric formula in Eq. (1).

**Community density**

In contrast to community likelihood, which quantifies the contribution of a community to the overall edge likelihood, community density simply measures the overall strength of connections within the community. By considering edge probabilities that are not conditioned on the community $C$, density implicitly takes into consideration potentially hierarchical and overlapping community structures. When a community is nested inside other communities, these enclosing communities contribute to the increased density of community's internal edges.

Formally, we define the density of a community as the joint probability of the edges between community members. Assuming network perturbation intensity $\alpha$, density of community $C$ is

defined as:

$$f_{\text{d}}(C|\alpha) = \prod_{\substack{(u,v)\in\mathcal{E} \\ u\in C, v\in C}} p(u,v|\alpha), \tag{3}$$

where $p(u,v|\alpha)$ is the probability of edge $(u,v)$ under network perturbation intensity $\alpha$. We derive analytical expression for $p(u,v|\alpha)$ which allows us to compute their values without ever actually perturbing the network (Supplementary Note 2).

**Community boundary**

To complement the internal connectivity measured by community density, community boundary considers the strength of edges leaving the community. A structural feature of a high quality community is its good separation from the surrounding parts of the network. In other words, a high quality community should have sharp edge boundary, *i.e.* $B_C = \{(u,v) \in \mathcal{E}; u \in C, v \notin C\}$[4]. This intuition is captured by accumulating the likelihood against edges connecting the community with the rest of the network:

$$f_{\text{b}}(C|\alpha) = \prod_{\substack{u\in C \\ v\in\mathcal{V}\setminus C}} (1 - p(u,v|\alpha)). \tag{4}$$

The evaluation of Eq. (4) takes computational time linear in the size of the network, which is impractical for large networks with many detected communities. To speed up the calculations, we use negative sampling (Supplementary Note 2) to calculate the value of Eq. (4), and thereby reduce the computational complexity of the boundary metric to time that depends linearly on the number of edges leaving the community.

**Community allegiance**

Last we introduce community allegiance. We define community allegiance as the preference for nodes to attach to other nodes that belong to the same community. Allegiance measures the fraction of nodes in a community for which the total probability of edges pointing inside the community is larger than probability of edges that point to the outside of the community. For a given community $C$ and network perturbation intensity $\alpha$, community allegiance is defined as:

$$f_{\text{a}}(C|\alpha) = \frac{1}{|C|} \sum_{u\in C} \delta\Big(\sum_{v\in N_u\cap C} p(u,v|\alpha) \geq \sum_{v\in N_u\setminus C} p(u,v|\alpha)\Big), \tag{5}$$

where $N_u$ is a set of network neighbors of $u$ and $\delta$ is the indicator function, $\delta(x) = 1$ if $x$ is true, and $\delta(x) = 0$, otherwise.

Community has high allegiance if nodes in the community tend to be more strongly connected to other members of the community than to the rest of the network. In a community with no significant allegiance this metric takes a value that is close to zero or changes substantially when the network is only slightly perturbed. However, in the presence of substantial community allegiance, the metric takes large values and is not sensitive to edge perturbation.

## Combining community prioritization metrics

We just defined four community prioritization metrics: likelihood, density, boundary, and allegiance. Each metric on its own provides a useful signal for prioritizing communities (Supplementary Note 8). However, scores of each metric might be biased, have high variance, and behave differently across different networks (Supplementary Figure 6). It is thus essential to combine the values of individual metrics into a single aggregated score.

We develop an iterative unsupervised rank aggregation method that, without requiring an external gold standard, combines the prioritization metrics into a single aggregated prioritization of communities. The method is outlined in Figure 5. It naturally takes into consideration the fact that importance of individual prioritization metrics varies across networks and across community detection methods. The aggregation method starts by representing the values of each prioritization metric with a ranked list. In each ranked list, communities are ordered by the decreasing value of the metric. The method then determines the contribution of each ranked list to the aggregate prioritization by calculating importance weights. The calculation is based on Bayes factors[36–38], an established tool in statistics. Each ranked list has associated a set of importance weights. Importance weights can vary with rank in the list. The method then calculates the aggregated prioritization of communities in an iterative manner by taking into account uncertainty that is present across different ranked lists and within each ranked list.

To calculate the weights without requiring gold standard, the method uses a two-stage iterative procedure. After initializing the aggregated prioritization, the method alternates between the following two stages until no changes in the aggregated prioritization are observed: (1) use the aggregated prioritization to calculate the importance weights for each ranked list, and (2) re-aggregate the ranked lists based on the importance weights calculated in the previous stage.

The model for aggregating community prioritization metrics, the algorithm, and the analysis of its computational time complexity are detailed in Supplementary Notes 4 and 5. The complete algorithm of CRank approach is provided in Supplementary Note 5.

**Code and data availability.** All relevant data are public and available from the authors of original publications. The project website is at: http://snap.stanford.edu/crank. The website contains preprocessed data used in the paper and additional examples of CRank's use. Source code of the CRank method is available for download from the project website.

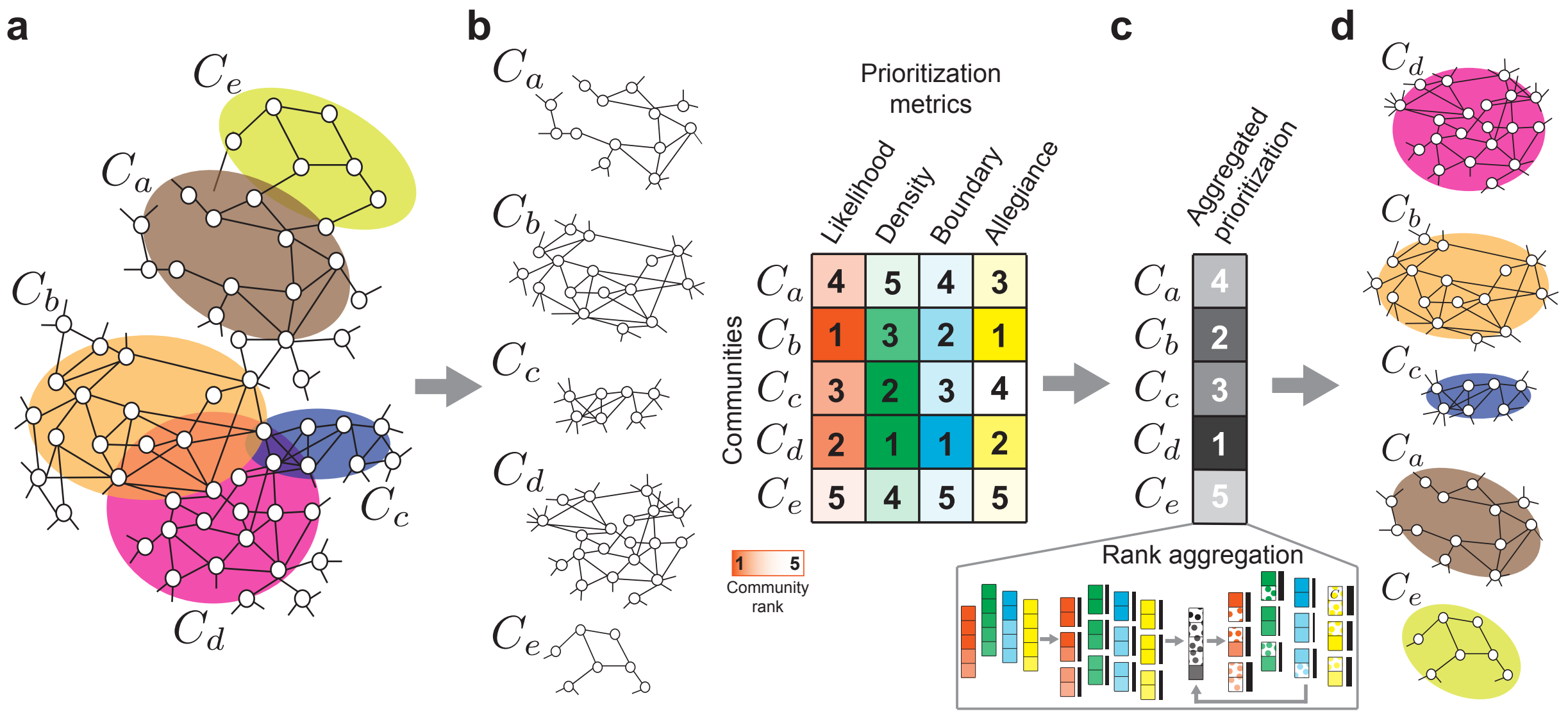

Figure 1: **Prioritizing network communities.** **(a)** Community detection methods take as input a network and output a grouping of nodes into communities. Highlighted are five communities, $(C_a, \ldots, C_e)$, that are detected in the illustrative network. **(b)** After communities are detected, the goal of community prioritization is to identify communities that are most promising targets for follow-up investigations. Promising targets are communities that are most associated with external network functions, such as cellular functions in protein-protein interaction networks, or cell types in cell-cell similarity networks. CRANK is a community prioritization approach that ranks the detected communities using only information captured by the network structure and does not require any external data about the nodes or edges of the network. However, when external information about communities is available, CRANK can make advantage of it to further improve performance (Supplementary Notes 9 and 10). CRANK starts by evaluating four different structural features of each community: the overall likelihood of the edges in the community (Likelihood), internal connectivity (Density), external connectivity (Boundary), and relationship with the rest of the network (Allegiance). CRANK can also integrate any number of additional user-defined metrics into the prioritization without any further changes to the method. **(c)** CRANK then applies a rank aggregation method to combine the metrics and **(d)** produce the final ranking of communities.

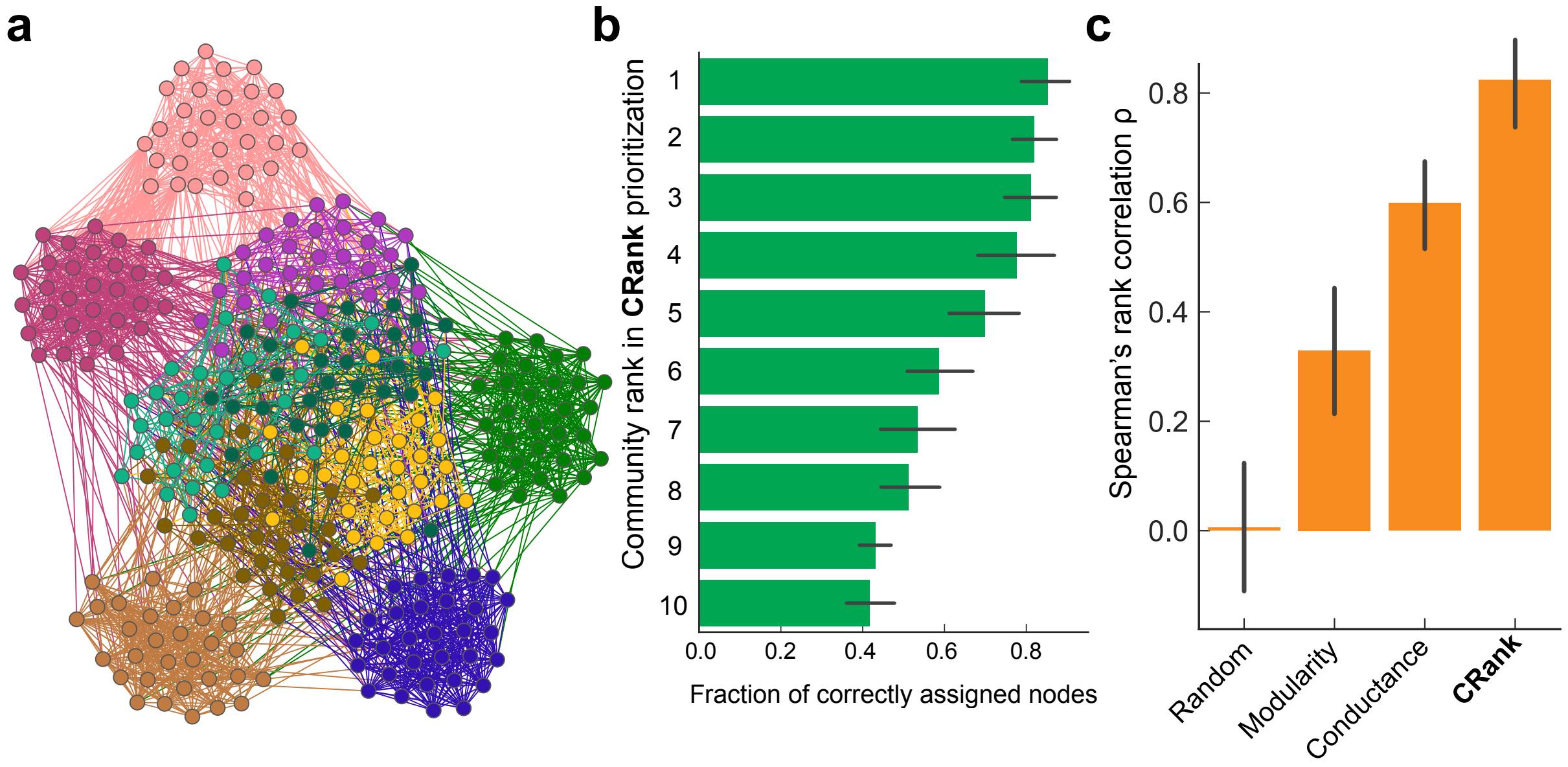

Figure 2: **Synthetic networks with planted community structure.** **(a)-(c)** In networks with known modular structure we can evaluate community prioritization by quantifying the correspondence between detected communities and the planted communities. **(a)** Benchmark networks on $N = 300$ nodes are created using a stochastic block model with 10 planted communities[10]. Each planted community has 30 nodes, which are colored by their planted community assignment. Planted communities use different values for within-community edge probability $p_{\text{in}}$, five use $p_{\text{in}} = 0.6$ and five use $p_{\text{in}} = 0.2$. As a result, planted communities with smaller within-community probability $p_{\text{in}}$ are harder to detect. For each benchmark network we apply a community detection method[6] to detect communities and then use CRANK to prioritize them. CRANK produces a ranked list of detected communities. The gold standard rank of each community is determined by how accurately it corresponds to its planted counterpart. **(b)** Each bar represents one detected community and the bars are ordered by CRANK's ranking with the highest ranked community located at the top and the lowest ranked community located at the bottom. As a form of validation, the width of each bar corresponds to the fraction of nodes in a community that are correctly classified into a corresponding planted community, with error bars showing the 95% confidence intervals over 500 benchmark networks. A perfect prioritization ranks the bars by decreasing width. Notice that CRANK perfectly prioritizes the communities even though it only uses information about the network structure, and has no access to information about the planted communities. **(c)** Prioritization performance is measured using Spearman's rank correlation $\rho$ between the generated ranking and the gold standard ranking of communities. A larger value of $\rho$ indicates a better performance. Across all benchmark networks, CRANK achieved average Spearman's rank correlation of $\rho = 0.82$. Alternative approaches resulted in poorer average performance: ranking based on modularity and conductance achieved $\rho = 0.33$ and $\rho = 0.60$, respectively, whereas random prioritization obtained $\rho = 0.00$.

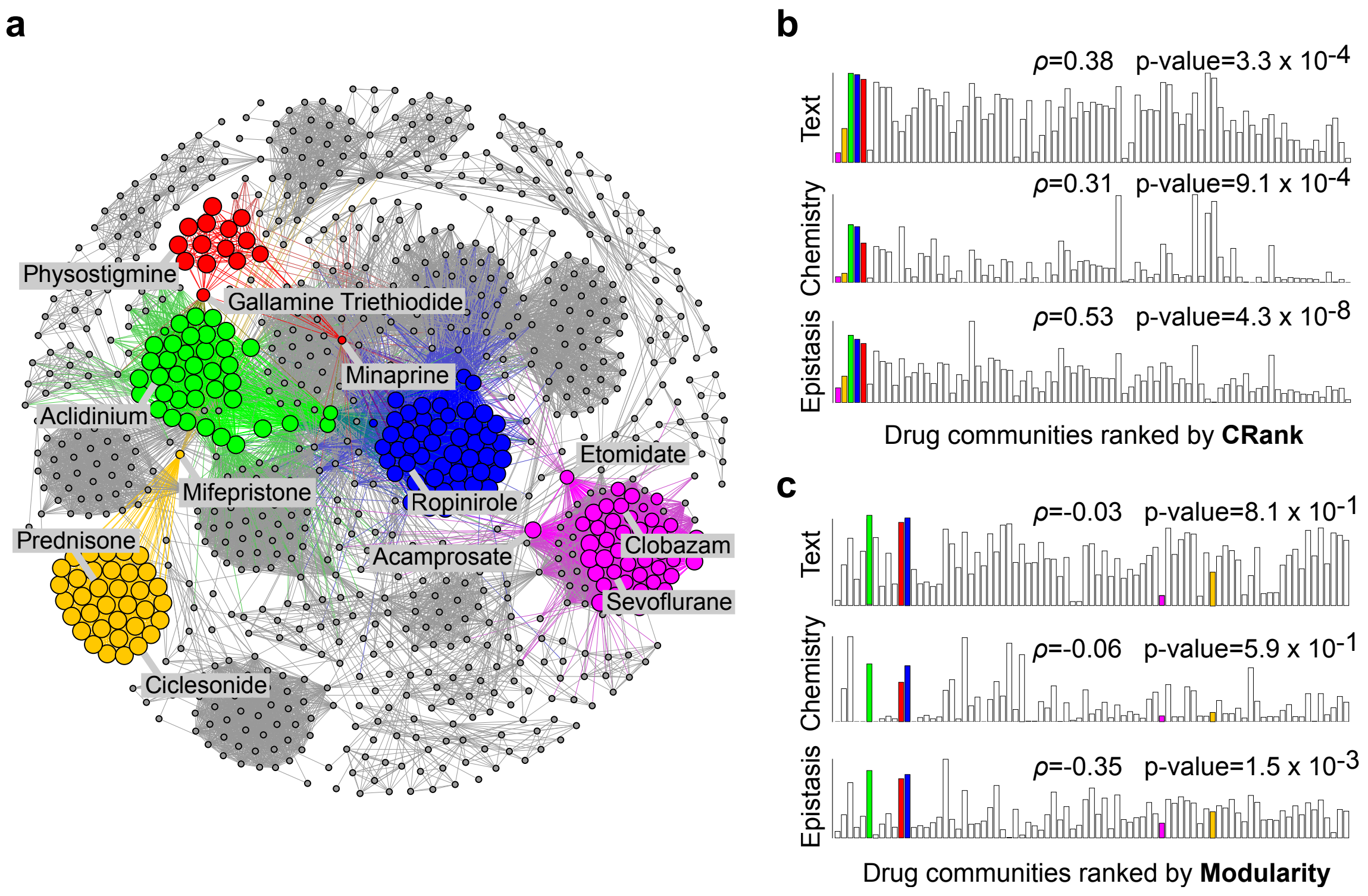

Figure 3: **Prioritizing network communities in the network of medical drugs.** **(a)** The network of medical drugs connects two drugs if they share at least one target protein. Communities were detected by a community detection method[7], and then prioritized by CRANK. Highlighted are five highest ranked communities as determined by CRANK. Nodes of the highlighted communities are sized by their score of the Likelihood prioritization metric (Supplementary Note 3). Investigation reveals that these communities contain drugs used to: treat asthma and allergies (*e.g.*, prednisone, ciclesonide; yellow nodes), induce anaesthesia or sedation (*e.g.*, clobazam, etomidate, sevoflurane, acamprosate; magenta nodes), block neurotransmitters in central and peripheral nervous systems (*e.g.*, physostigmine, minaprine, gallamine triethiodide; red nodes), block the activity of muscarinic receptors (*e.g.*, acidinium; green nodes), and activate dopamine receptors (*e.g.*, ropinirole; blue nodes). **(b-c)** We evaluate community prioritization against three external chemical databases (Supplementary Note 6) that were not used during community detection or prioritization. For each community we measure: (1) drug-drug interactions between the drugs ("Epistasis"), (2) chemical structure similarity of the drugs ("Chemistry"), and (3) associations between drugs derived from text data ("Text"). We expect that a true high-priority community will have more drug-drug interactions, higher similarity of chemical structure, and stronger textual associations between the drugs it contains. Taking this into consideration, the external chemical databases define three gold standard rankings of communities against which CRANK is evaluated. Bars represent communities; bar height denotes similarity of drugs in a community with regard to the gold standard based on external chemical databases. In a perfect prioritization, bars would be ordered such that the heights would decrease from left to right. **(b)** CRANK ranking of drug communities outperforms ranking by modularity **(c)** across all three chemical databases (as measured by Spearman's rank correlation $\rho$ with the gold standard ranking). CRANK ranking achieves $\rho = 0.38, 0.31, 0.53$, while modularity obtains $\rho = -0.03, -0.06, -0.35$.

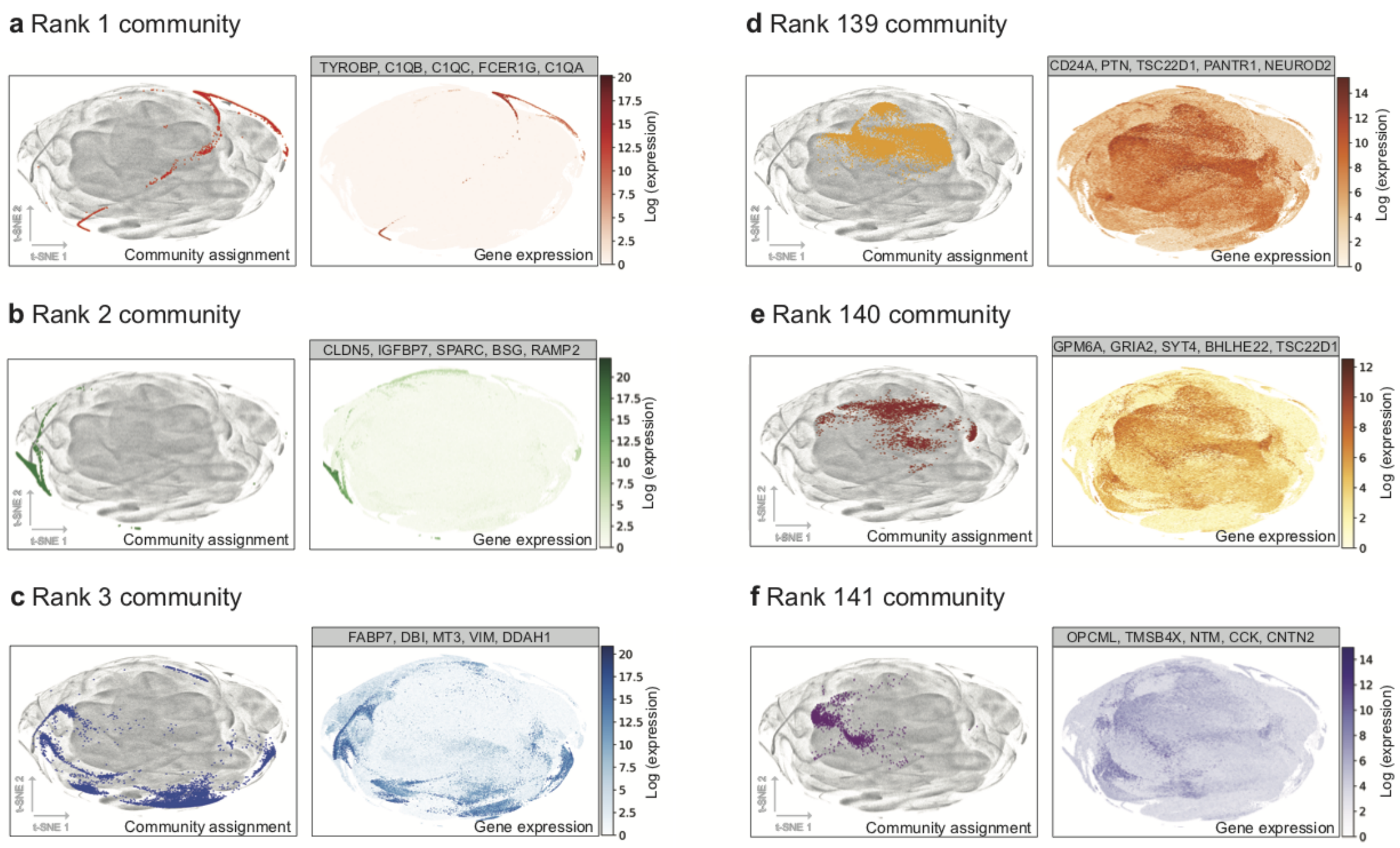

Figure 4: **Prioritizing network communities in the megascale cell-cell similarity network.** The network of embryonic mouse brain has 1,306,127 nodes representing brain cells[29]. Communities are detected using a community detection method developed for single-cell RNA-seq data[8] and prioritized using CRANK, generating a rank-ordered list of detected communities. **(a-c)** Shown are three communities that are ranked high by CRANK; **(a)** rank 1, **(b)** rank 2, and **(c)** rank 3 community. t-SNE projections[39] show cells assigned to each community. t-SNE is a dimensionality reduction technique that is particularly well suited for visualization of high-dimensional data. Cells assigned to each community are distinguished by color, and all other cells are shown in grey. We investigate the quality of community ranking by examining gene markers for cells in each community[28]. We use the single-cell RNA-seq dataset to obtain a gene expression profile for each cell, indicating the activity of genes in the cell. For each community we then identify marker genes, *i.e.*, genes with the strongest differential expression between cells assigned to the community and all other cells[9]. In the t-SNE projection we then color the cells by how active the marker genes are. This investigation reveals that communities ranked high by CRANK are represented by clusters of cells whose marker genes have a highly localized expression. For example, marker genes for rank 1 community in (a) (the highest community in CRANK ranking) are *TYROBP*, *C1QB*, *C1QC*, *FCER1G* and *C1QA*. Expression of these genes is concentrated in cells that belong to the rank 1 community. Similarly, marker genes for rank 2 and rank 3 communities are specifically active in cell populations that match well the boundary of each community. **(d-f)** t-SNE projections show cells assigned to 3 low-ranked communities; **(d)** rank 139, **(e)** rank 140, and **(f)** rank 141 community. t-SNE projections are produced using the same differential analysis as in (a-c). Although these communities correspond to clusters of cells in the t-SNE projections, their marker genes have diluted gene expression that is spread out over the entire network, indicating that CRANK has correctly considered these communities to be low priority. For example, marker genes for rank 141 community in (f) are *OPCML*, *TMSB4X*, *NYM*, *CCK*, and *CNTN2*, which show a weak expression pattern that is diffused across the entire network.

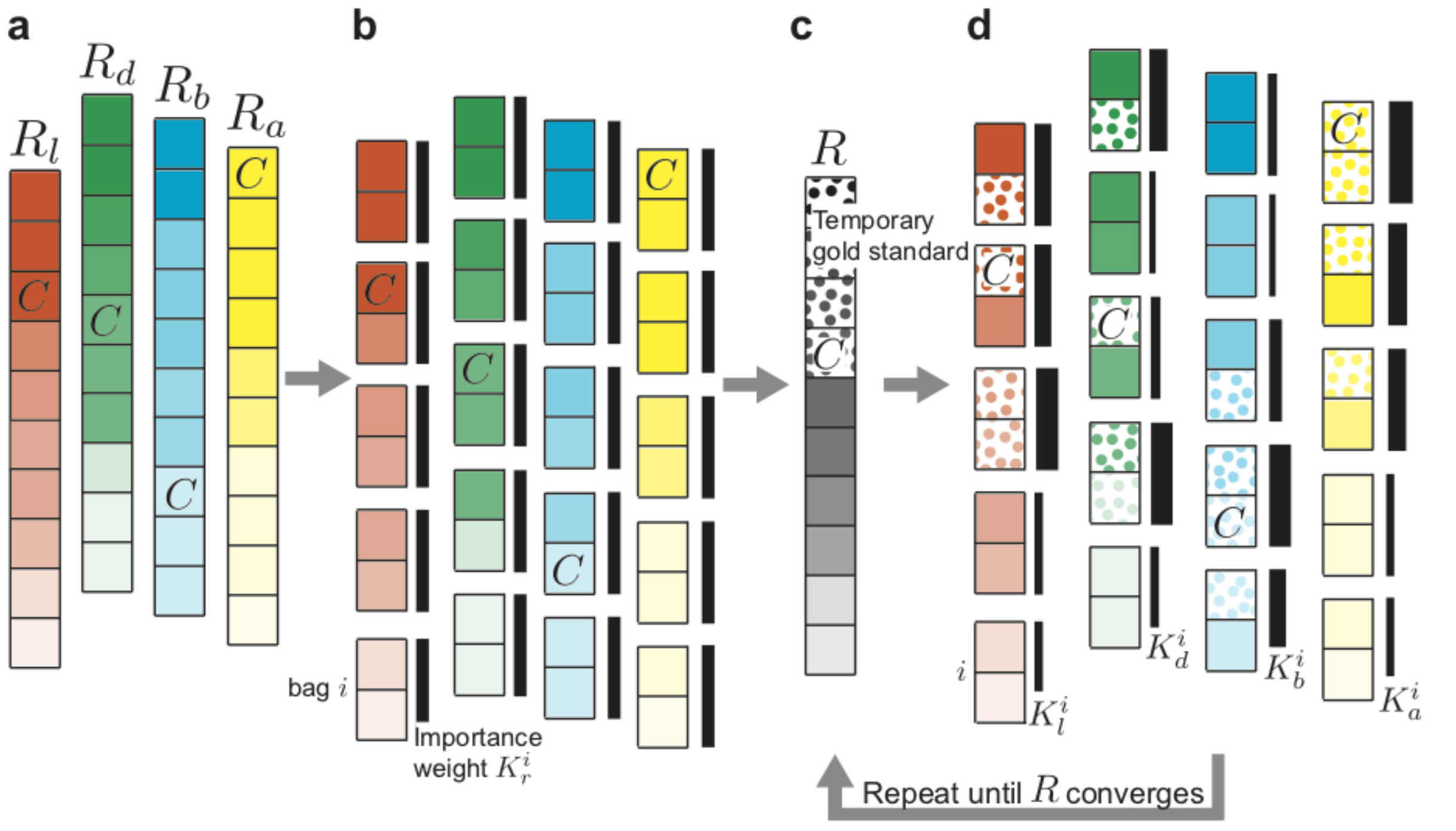

Figure 5: **Combining community prioritization metrics without an external gold standard. (a)** The rank aggregation algorithm starts with four ranked lists of communities, $R_r$, each one arising from the values of a different community prioritization metric $r$ (where $r$ is one of "l" – likelihood, "d" – density, "b" – boundary, "a" – allegiance). Communities are ordered by the decreasing value of the metric. We use $C$ to indicate the rank of an illustrative community by the community prioritization metrics and at different stages of the algorithm. **(b)** Each ranked list is partitioned into equally sized groups, called bags. Each bag $i$ in ranked list $R_r$ has attached importance weight $K_r^i$ whose initial values are all equal (represented by black bars all of same width). CRANK uses the importance weights $K_r^i$ to initialize aggregate prioritization $R$ as a weighted average of community ranks $R_{\mathrm{l}}$, $R_{\mathrm{d}}$, $R_{\mathrm{b}}$, $R_{\mathrm{a}}$. **(c)** The top ranked communities (denoted as dotted cells) in the aggregated prioritization $R$ serve as a temporary gold standard, which is then used to iteratively update the importance weights $K_r^i$. **(d)** In each iteration, CRANK updates importance weights using the Bayes factor calculation[36] (Supplementary Note 4). Given bag $i$ and ranked list $R_r$, CRANK updates importance weight $K_r^i$, based on how many communities from the temporary gold standard appear in bag $i$. Updated importance weights then revise the aggregated prioritization in which the new rank $R(C)$ of community $C$ is expressed as: $R(C) = \sum_r \log K_r^{i_r(C)} R_r(C)$, where $K_r^{i_r(C)}$ indicates the importance weight of bag $i_r(C)$ of community $C$ for metric $r$, and $R_r(C)$ is the rank of $C$ according to $r$. By using an iterative approach, CRANK allows for the importance of a metric not to be predetermined and to vary across communities.

**Acknowledgements.** M.Z., R.S., and J.L. were supported by NSF, NIH BD2K, DARPA SIMPLEX, Stanford Data Science Initiative, and Chan Zuckerberg Biohub.

**Author contribution.** M.Z., R.S. and J.L. designed and performed research, contributed new analytic tools, analyzed data, and wrote the paper.

**Author information.** The authors declare no conflict of interest. Correspondence should be addressed to J.L. (jure@cs.stanford.edu).

**Additional information.** Supplementary Notes contain a detailed description of the community prioritization approach, descriptions of datasets, experimental setup, and additional experiments. Supplementary Table 1 contains detailed community prioritization results for the medical drug network. Code to run CRANK and examples are at: http://snap.stanford.edu/crank.

Supplementary materials for

# Prioritizing network communities

Marinka Zitnik,[1] Rok Sosič,[1] Jure Leskovec[1,2,*]

[1]Computer Science Department, Stanford University

[2]Chan Zuckerberg Biohub, San Francisco, CA

[*]Corresponding author. E-mail: jure@cs.stanford.edu

This PDF file includes:

- Supplementary Notes 1 to 11
- Supplementary Figures 1 to 8
- Supplementary Tables 1 to 9
- References

Other supporting material for this manuscript includes the following:

- Supplementary Table with prioritization results for the medical drug network

## Supplementary Note 1 Document outline

In this document, we present a detailed description of the community prioritization approach, discussion of the datasets used and their analysis. First, we describe a network perturbation model used by CRANK and then derive expressions for edge probabilities in this model (Supplementary Note 2). The derived expressions enable us to estimate edge probabilities in a perturbed network in a closed form manner. These estimates are essential components of CRANK community prioritization metrics. We then provide details on computing the metrics, beyond those presented in the main text (Supplementary Note 3). We proceed by describing CRANK rank aggregation method (Supplementary Note 4). Its role is to combine the metric scores and form an aggregated prioritization of communities. We then provide a detailed description of complete CRANK approach (Supplementary Note 5).

We describe network data used in experiments (Supplementary Note 6). We outline experimental setup, overview community detection methods considered in the paper, and describe alternative techniques for community prioritization and for rank aggregation (Supplementary Note 7).

Finally, we present further results of empirical evaluations. In Supplementary Note 8 we report additional experiments on real-world networks, and we further investigate CRANK's properties. In Supplementary Note 9 we show how to integrate any number of additional user-defined metrics into CRANK without requiring further technical changes to the CRANK model. In Supplementary Note 10 we show how CRANK can use domain-specific or other meta and label information to supervise community prioritization. In Supplementary Note 11 we describe additional experiments on medical, social, and information networks, beyond those presented in the main text.

## Supplementary Note 2 Network perturbation model

Our goal in this note is to find closed form expressions that will enable us to analytically quantify how stable are communities if the network is perturbed. These expressions are important because they allow us to avoid instantiating any of the perturbed networks when computing community prioritization metrics. Consequently, CRANK easily scales to large networks.

Notice that our ability to analytically compute perturbation effects offers significant improvement over established methods, such as, for example, methods for evaluating the quality of network community structure[1–6]. Methods of this kind explicitly perturb the network many times. They evaluate the quality of community structure by partitioning an entire network, applying the network rewiring model many times, materializing hundreds of perturbed networks and then running community detection repeatedly on all perturbed network versions. Such methods, however, can suffer from expensive computation and are computationally prohibitive for large networks. Details are provided next.

### Supplementary Note 2.1 Network perturbation

We start by describing a network perturbation model that can perturb an arbitrary network by an arbitrary amount based on network's node degree distribution. To formulate the probabilities of edges potentially arising when perturbing an arbitrary network by an arbitrary amount we consider a network rewiring model. We restrict our perturbed networks to have the same number of nodes and edges as the original unperturbed network, only edges are randomly rewired. We measure perturbation intensity by a parameter $\alpha$, where a value of $\alpha$ close to zero indicates that a network is perturbed by only a small amount and has only a few edges rewired. Perturbation intensity close to one corresponds to a perturbed network, which is almost completely random and uncorrelated with the original network.

Given a network $G(\mathcal{V}, \mathcal{E})$, whose nodes are given by $\mathcal{V}$ and edges by $\mathcal{E}$, we denote the network resulting from $\alpha$-perturbing edges in $G$ as: $G(\alpha) = G(\mathcal{V}, \mathcal{E}(\alpha))$, $0 \leq \alpha \leq 1$. $\alpha$ denotes perturbation intensity. This means that $G(0)$ (*i.e.*, $\alpha = 0$) is identical to the original network, $G(0) = G$, since no edge has changed its position in the network, whereas $G(1)$ (*i.e.*, $\alpha = 1$) is a maximally perturbed network obtained by rewiring all edges in $G$ such that node degree distribution of $G$ is preserved in $G(1)$.

Given $\alpha$, we specify the network $G(\alpha)$ by perturbing the network $G$ as follows[2]. We consider each edge $(u, v) \in \mathcal{E}$ in network $G$ in turn and either:

- with probability $\alpha$ we add an edge $(u', v')$ to $G(\alpha)$ such that the probability of edge falling between nodes $u'$ and $v'$ is $e_{u'v'}/m$, or
- with probability $1 - \alpha$ we add an edge $(u, v)$ to $G(\alpha)$.

Here, $e_{u'v'} = k_{u'}k_{v'}/(2m)$, where $k_{u'}$ is the degree of node $u'$ in $G$, denoted also as $k_{u'} = |N_{u'}|$, and $m$ is the number of edges in network $G$, $m = |\mathcal{E}|$. This network rewiring model generates networks $G(\alpha)$ that not only have the same number of edges as the original network $G$, but in which the expected degrees of nodes are the same as the original degrees[2].

### Supplementary Note 2.2 Statistical community detection model

Let us suppose we are given a network $G(\mathcal{V}, \mathcal{E})$, and a community detection model $M$ that detects communities $\mathcal{C}$, $\mathcal{C} = \{C; C \subseteq \mathcal{V}\}$, in network $G$. Here, every community $C$ is given by a set of its member nodes.

We assume that $M$ is a statistical community detection model (*e.g.*,[7–19]). In that case, $M$ allows us to evaluate: (1) the probability of node $u$ belonging to a community $C$, $p_C(u) = p(u \in C)$, (2) the probability of an edge, $p(u, v) = p((u, v) \in \mathcal{E})$, and (3) the probability of an edge from node $u$ to node $v$ conditioned on nodes' joint affiliation with a community $C$. We denote the latter probability as $p_C(u, v) = p((u, v) \in \mathcal{E} | u \in C, v \in C)$ and view it as a contribution of community $C$ towards the creation of edge $(u, v)$.

Commonly used community detection methods, like the Stochastic Block Model[7,10,16,20,21], Affiliation Graph Model[8,9], Latent Feature Graph Model[11–15], and Attributed Graph Model[17–19] all allow for computing the above three quantities.

Next, we use the quantities (1)–(3) to specify edge probabilities and node-community affiliation probabilities arising under the network perturbation model from Supplementary Note 2.1.

### Supplementary Note 2.3 Edge probabilities in perturbed network

We express the probability of an edge $(u, v)$ appearing in a perturbed network $G(\alpha)$ as a function of the probability of edge $(u, v)$ appearing in the original network $G$ and of perturbation intensity $\alpha$. The expressed probability is denoted as $p(u, v|\alpha)$.

There are two ways by which nodes $u$ and $v$ can be connected with an edge $(u, v)$ in the perturbed network $G(\alpha)$. If an edge $(u, v)$ exists in $G$, then with probability $1 - \alpha$ the edge is retained during perturbation. Otherwise, nodes $u$ and $v$ can connect in $G(\alpha)$ as a result of network rewiring as described in Supplementary Note 2.1. In the latter case, edge $(u, v)$ appears in $G(\alpha)$ if it is a replacement for any of the expected $\alpha m$ edges that change their original positions in network $G$. This reasoning gives us the probability of edge $(u, v)$ emerging in perturbed network $G(\alpha)$ as:

$$p(u, v|\alpha) = p(u, v)(1 - \alpha) + (1 - p(u, v))(1 - (1 - \frac{e_{uv}}{m})^{\alpha m}), \tag{1}$$

where $e_{uv}$ is equal to $e_{uv} = k_u k_v/(2m)$. Notice that expression in Eq. (1) approximates probability of an edge in a perturbed network. This is because it considers the expected fraction of rewired edges in a perturbed network, but it ignores variance and skewness of rewiring distribution. We empirically validated the expression by comparing it with results obtained by explicitly perturbing the network many times. We observed that analytical expression for the edge probability in Eq. (1) led to an accurate estimation of empirical results for most considered real-world networks.

An approach, analogous to the derivation of probability $p(u, v|\alpha)$, also gives us the probability that a community $C$ detected in the original network $G$ generates a particular edge in the perturbed network $G(\alpha)$. Probability $p_C(u, v|\alpha)$ that an edge $(u, v)$ whose both endpoints belong to community $C$ is included in the perturbed network can be written as:

$$p_C(u,v|\alpha) = p_C(u,v)(1-\alpha) + (1-p_C(u,v))(1-(1-\frac{e_{uv}}{m})^{\alpha m}). \tag{2}$$

We use expressions in Eq. (1) and Eq. (2) to specify the probability of an edge $(u,v)$ whose endpoints belong to community $C$ as:

$$s_C^{(1)}(u,v|\alpha) = p((u,v) \in \mathcal{E}(\alpha), u \in C, v \in C) = p_C(u)p_C(v)p_C(u,v|\alpha). \tag{3}$$

Likewise, the probability of a non-edge between nodes $u$ and $v$ that are both assigned to a community $C$ is equal to:

$$s_C^{(2)}(u,v|\alpha) = p((u,v) \notin \mathcal{E}(\alpha), u \in C, v \in C) = p_C(u)p_C(v)(1-p_C(u,v|\alpha)). \tag{4}$$

Recall that $\alpha$ measures the intensity of network perturbation. By varying the value for $\alpha$, the intensity of network perturbation is interpolated between two extreme cases:

- $\alpha = 1$ corresponds to a perturbation that generates a network $G(1)$, whose edge probabilities as returned by Eq. (1–2) are completely determined by the perturbation model.
- $\alpha = 0$ corresponds to a perturbation the regenerates the original network, $G(0) = G$, meaning that edge probabilities as returned by Eq. (1–2) are exactly the same as in the original network, *e.g.*, $p(u,v|\alpha = 0) = p(u,v)$.

## Supplementary Note 3 Structural features of network communities

CRANK prioritizes communities based on the robustness and magnitude of multiple structural features of each community. CRANK defines community prioritization metrics, which capture key structural features and characterize network connectivity for each community. In this note we provide further details on two metrics, beyond those presented in the main text.

### Supplementary Note 3.1 Further details on computing community likelihood

The main text defines community likelihood that is calculated for each community. We also define likelihood score of every node in a given community. Likelihood score of a node $u$ in a community $C$ is the product of community-dependent probabilities of both edges and non-edges adjacent to node $u$:

$$n_l(u|C,\alpha) = p_C(u) \prod_{v \in C} s_C(u,v|\alpha) \tag{5}$$

where $s_C(u, v|\alpha)$ is defined as follows:

$$s_C(u, v|\alpha) = \begin{cases} s_C^{(1)}(u, v|\alpha)/p_C(u) \text{ if } (u, v) \in \mathcal{E} \\ s_C^{(2)}(u, v|\alpha)/p_C(u) \text{ if } (u, v) \notin \mathcal{E}, \end{cases}$$

where $s_C^{(1)}$ and $s_C^{(2)}$ are defined in Eq. (3) and Eq. (4), respectively. The node likelihood formula $n_l$ gives us an alternative way to express community likelihood. That is, likelihood of community $C$, $f_l(C|\alpha)$, can be seen as a product of likelihood scores for all the nodes that are affiliated with $C$: $f_l(C|\alpha) = \prod_{u \in C} n_l(u|C, \alpha)$.

**Supplementary Note 3.2 Further details on computing community boundary**

The evaluation of formula for community boundary $f_b(C|\alpha)$ takes computational time linear in the size of the network, which is impractical for large networks with many detected communities. To speed up the calculations, we use negative sampling to calculate the value of $f_b(C|\alpha)$, and thereby reduce the computational complexity of the boundary metric to time that depends linearly on the number of edges leaving the community.

We use negative sampling[22–24] to obtain a computationally efficient approximation of community boundary. In general, negative sampling can be used to approximate a function whose evaluation takes into consideration the entire universe of objects in a domain, such as all nodes in a large network. We calculate community boundary $f_b$ using the negative sampling as:

$$f_b(C|\alpha) = \prod_{\substack{u \in C \\ v \in \mathcal{V} \setminus C, (u,v) \in \mathcal{E}}} (1 - p(u, v|\alpha)) \prod_{\substack{u \in C \\ i=1}}^{k} (1 - p(u, v_i|\alpha)), \quad (6)$$

where $P_C$ is a noise distribution from which $k$ nodes $v_i$ are drawn, $v_i \sim P_C$. Formula Eq. (6) is used to replace community boundary formula given in the main text. Noise distribution $P_C$ is a uniform distribution defined over the non-edge boundary $T_C = \{v; v \in \mathcal{V} \setminus C, \nexists u \in C : (u, v) \in \mathcal{E}\}$. Formula $f_b$ considers $k$ non-edges for each node in community $C$. Our experiments indicate that values of $k$ in the range 5–20 are useful for small networks, while for large networks the value of $k$ can be as small as 2–5. This observation is aligned with the previous work in negative sampling[22–24].

In other words, Eq. (6) says that when computing community boundary $f_b$, for a given node $u \in C$, we consider all nodes that lie outside of $C$ but are connected with node $u$ (*i.e.*, first product term in Eq. (6)), and also randomly selected nodes that are neither assigned to $C$ nor linked with $u$ (*i.e.*, second product term in Eq. (6)). The latter nodes are selected uniformly at random from the set $T_C$. This formulation posits that a high quality community should have sharp edge boundary[25].

Importantly, the formula in Eq. (6) allows us to reduce computational time complexity of community boundary for a given node $u$ from being proportional to the number of nodes in the network (*i.e.*, $|\mathcal{V}|$) to being proportional to the size of community $C$ plus the number of random non-edges (*i.e.*, $|C|+k$). Since communities in real-world settings are much smaller than the entire network, negative sampling allows us to scale community boundary to large networks.

## Supplementary Note 4 Rank aggregation model

Prioritization of communities involves measuring different network structural features of communities. The features are measured by four prioritization metrics: community likelihood, community density, community boundary and community allegiance. Next, we describe how to combine the scores of different metrics into an aggregated prioritization of communities.

The simplest approach to combining the metrics is to treat all the metrics equally and average their scores. While such an approach does not need any external gold standard ranking of communities it can be unacceptably sensitive to noise and outliers (Supplementary Note 8.4). One alternative is to evaluate individual metrics against an external gold standard ranking. However, we need to examine all the communities and rank them in order to obtain the gold standard, which is precisely the task we try to avoid.

We adopt a statistical approach and propose a rank aggregation method that combines the scores of different metrics. Furthermore, our proposed rank aggregation method operates without requiring a gold standard ranking of communities. Details are provided next.

### Supplementary Note 4.1 Ranked lists of communities

The rank aggregation method starts with four ranked lists, one from each of the four prioritization metrics, where communities are ordered by their scores such that communities with the highest score are at the beginning of each list. The rank of a community is its position in the list. Given the scores $r_f$ for a network structural feature $f$, the ranked list $R_r$ is:

$$R_r(C) = 1 + \sum_{D \neq C} I(r_f(C;\alpha) \leq r_f(D;\alpha)) \qquad \text{for } C \in \mathcal{C}, \tag{7}$$

signifying that $R_r(C)$ is the rank of community $C$ according to the scores of metric $r$. The function $I$ is the indicator function, such that $I(X) = 1$ if $X$ is true and $I(X) = 0$ otherwise, and $\mathcal{C}$ is the set of all communities found in a given network by a given community detection method. To assign ranks to communities with tied scores, we consider the average of the ranks that would have been assigned to all the tied communities and assign this average to each community.

Given the ranked lists $R_l$, $R_d$, $R_b$ and $R_a$, we wish to combine the ranked lists into a single ranked list $R$. The ranked list $R$ *is* CRANK*'s final result representing the aggregated prioritization of communities.*

### Supplementary Note 4.2 Background on Bayes factors

We proceed by describing the Bayes factors, a tool in statistics[26–30], that is the centerpiece of our method for combining prioritization metrics. We use Bayes factors to estimate the weights to be attached to the ranked lists of communities so that we can obtain the aggregated prioritization of communities that takes account of uncertainty present in the ranked lists arising from different prioritization metrics.

#### Supplementary Note 4.2.1 The Bayes factor of one ranked list

We begin with a single ranked list $R_r$, assumed to have arisen under one of the two hypotheses $H_r^1$ and $H_r^2$ according to probability density $p(R_r|H_r^1)$ and $p(R_r|H_r^2)$, respectively. Using the Bayesian formulation[30], the two hypotheses are:

$$H_r^1 \quad - \quad \text{Scores of metric } r \text{ match the gold standard } R^*, \tag{8}$$

$$H_r^2 \quad - \quad \text{Scores of metric } r \text{ do not match the gold standard } R^*. \tag{9}$$

Here, $R^*$ is a ranked list representing the gold standard ranking of communities. For now, we assume that the gold standard $R^*$ is given, we will later in Supplementary Note 4.4 discuss how to determine probability densities $p(R_r|H_r^1)$ and $p(R_r|H_r^2)$ when the gold standard $R^*$ is not available.

Given prior probabilities $p(H_r^1)$ and $p(H_r^2) = 1 - p(H_r^1)$, the ranked list $R_r$ produces posterior probabilities $p(H_r^1|R_r)$ and $p(H_r^2|R_r) = 1 - p(H_r^1|R_r)$. The posterior probability can be related to the prior probability using the Bayes' theorem as:

$$p(H_r^i|R_r) = \frac{p(R_r|H_r^i)p(H_r^i)}{p(R_r|H_r^1)p(H_r^1) + p(R_r|H_r^2)p(H_r^2)}. \qquad (i = 1, 2) \tag{10}$$

In the odds scale[30] (odds = probability / (1 - probability)), the relation of posterior probability to prior probability takes the following form:

$$\frac{p(H_r^1|R_r)}{p(H_r^2|R_r)} = \frac{p(R_r|H_r^1)}{p(R_r|H_r^2)} \frac{p(H_r^1)}{p(H_r^2)}. \tag{11}$$

This means that transformation of the prior odds to the posterior odds involves multiplication by a factor:

$$K_r = \frac{p(R_r|H_r^1)}{p(R_r|H_r^2)}, \tag{12}$$

which is known as the Bayes factor[26,28,31,32] for comparing hypotheses $H_r^1$ and $H_r^2$. Thus, in words, Eq. (11) is equal to:

$$\text{posterior odds} = \text{Bayes factor} \times \text{prior odds}, \tag{13}$$

which means that the Bayes factor $K_r$ can also be written as the ratio of the posterior odds to the prior odds:

$$K_r = \frac{p(H_r^1|R_r)}{p(H_r^2|R_r)} \frac{p(H_r^2)}{p(H_r^1)}, \tag{14}$$

and can be used to quantify the evidence[26] provided by ranked list $R_r$ in favor of hypothesis $H_r^1$. We use the Bayes factor $K_r$ to measure the relative success of $H_r^1$ and $H_r^2$ at predicting the gold standard ranking $R^*$: a Bayes factor greater than 1 means that the ranked list $R_r$ provides greater evidence for $H_r^1$, whereas a Bayes factor less than 1 means that the ranked list $R_r$ provides greater evidence for $H_r^2$.

### Supplementary Note 4.3 Aggregating ranked lists

We adopt a statistical approach to combine the ranked lists arising from different prioritization metrics. The approach specifies the Bayes factor for each ranked list following the exposition in Supplementary Note 4.2.1.

When several metrics are considered, the Bayes factors are obtained as follows. Given ranked lists $R_l$, $R_d$, $R_b$ and $R_a$, we consider pairs of hypotheses $(H_l^1, H_l^2)$, $(H_d^1, H_d^2)$, $(H_b^1,\ H_b^2)$, and $(H_a^1, H_a^2)$. The meaning of a hypothesis pair for metric $r$ is described in Eq. (8) and Eq. (9). We compare each of $H_l^1, H_d^1, H_b^1, H_a^1$ in turn with the corresponding hypothesis $H_l^2$, $H_d^2$, $H_b^2$, $H_a^2$ as described in Supplementary Note 4.2.1. Using the formula in Eq. (12), this procedure yields the Bayes factors $K_l, K_d, K_b$ and $K_a$. Following Eq. (10), we then calculate the posterior probability of $H_r^1$, *i.e.*, the posterior probability that ranked list $R_r$ matches the gold standard $R^*$, as:

$$p(H_r^1|R_l, R_d, R_b, R_a) = \frac{\alpha_r K_r}{\sum_{r'} \alpha_{r'} K_{r'}}, \tag{15}$$

where $\alpha_r = p(H_r^1)/p(H_r^2)$ is the prior odds for $H_r^1$ against $H_r^2$, and $r'$ goes over all considered metrics, $r' \in \{l, d, b, a\}$. In this paper, we take all the prior odds $\alpha_r$ equal to 1. Although this is a natural choice, we note that other values of $\alpha_r$ may be used to reflect prior information about the relative plausibility of different ranked lists.

The probabilities given by Eq. (15) lead directly to the prediction that takes account of uncertainty in the metrics[26,30]. Recall that we want to aggregate ranked lists $R_l$, $R_d$, $R_b$, $R_a$ into a single ranked list $R$ representing the aggregated prioritization of communities, *i.e.*, the final prediction of CRANK. This means we would like to calculate the probability of the aggregated prioritization $R$ conditioned on the information provided by the ranked lists. This probability can be written as:

$$p(R|R_l, R_d, R_b, R_a) = \sum_{r'} p(R|R_l, R_d, R_b, R_a, H^1_{r'})p(H^1_{r'}|R_l, R_d, R_b, R_a), \tag{16}$$

where we account for uncertainty by weighting each ranked list by how well it matches the gold standard $R^*$. We specify the posterior probability $p(H^1_{r'}|R_l, R_d, R_b, R_a)$ using the Bayes factor from Eq. (15). Finally, combining Eq. (15) and Eq. (16), we can write the probability for aggregated prioritization $R$ as:

$$p(R|R_l, R_d, R_b, R_a) = \frac{\sum_{r'} K_{r'}\, p(R|R_l, R_d, R_b, R_a, H^1_{r'})}{\sum_{r'} K_{r'}}. \tag{17}$$

The posterior probabilities expressed through Bayes factors favor those ranked lists that better match the gold standard $R^*$.

Examining Eq. (17), we see that ranked lists are aggregated as a weighted average with weights being equal to the Bayes factors of the ranked lists, *i.e.*, weight for ranked list $R_r$ is equal to $K_r/(\sum_{r'} K_{r'})$. This means that the aggregated prioritization $R$ is a weighted average of the ranked lists $R_l$, $R_d$, $R_b$, and $R_a$:

$$R = \frac{\sum_{r'} K_{r'} R_{r'}}{\sum_{r'} K_{r'}}. \tag{18}$$

In the next section, we describe how to determine the aggregated prioritization when the gold standard $R^*$ is not available, and how to learn the weights for each ranked list that vary with rank (*i.e.*, position) in the list.

### Supplementary Note 4.4 Estimating importance weights

We proceed by explaining how to estimate in practice the Bayes factors needed to calculate the aggregated prioritization. For that, we introduce *importance weights*, which follow directly from the Bayes factors described above.

#### Supplementary Note 4.4.1 Lack of gold standard community ranking

In order to aggregate the ranked lists of communities we need to calculate the Bayes factors that appear in the aggregated prioritization formula in Eq. (17) and Eq. (18). The evaluation of the Bayes factors entails computing the posterior probability for each ranked list. As we explain next, the calculation of the posterior probability requires a priori knowledge, which is practically impossible to obtain.

The Bayes factor $K_r$ of ranked list $R_r$ is defined as the evidence provided by the ranked list $R_r$ in favor of the gold standard $R^*$ (see Supplementary Note 4.2). Here, the gold standard $R^*$ is a ranked list that orders communities found in a network in the decreasing order of their importance for further investigation in the follow-up studies. Intuitively, community that ranks higher in $R^*$ should be better at representing a structure that carries a meaning in a given network (*e.g.*, a disease causing pathway of proteins in a protein-protein interaction network, or, a group of functionally similar products in the Amazon product co-purchasing network) than community that ranks lower in $R^*$.

However, it is practically impossible to a priori know which detected communities rank at the top of the gold standard $R^*$. In order to obtain such a gold standard ranking of communities, we need to examine all the communities by performing potentially costly and time consuming experiments. These experiments would allow us to determine for each community whether it corresponds to a meaningful network structure. Afterwards, we would construct the gold standard $R^*$ by ranking the communities based on the outcomes of the experiments. These experiments may render construction of the gold standard $R^*$ difficult or even impossible in practice. Furthermore, as the aim of community prioritization is to avoid the need to perform all the experiments, the ability to prioritize communities should not depend on the availability of $R^*$. We therefore resort to a different approach.

#### Supplementary Note 4.4.2 Bootstrapping the importance weights

We describe an approach that resolves the problem of aggregating ranked lists when the gold standard community ranking is not available. The approach takes as its input the ranked lists and it estimates the weights (*i.e.*, Bayes factors $K_r$) for each ranked list, which vary with rank in the list, in an unsupervised manner. This means the approach does not require communities with top aggregated ranks to be known a priori.

The approach uses a two-stage bootstrapping process to estimate the weights for each ranked list. This is achieved based on the ranked list decomposition rather than based on a gold standard community ranking. Details are provided next.

**Decomposing ranked lists into bags.** Each ranked list is partitioned into equally sized groups of communities that we call bags. Formally, bags correspond to sets of communities. The ranked list $R_r$ is partitioned into $B$ bags. The $j$-th bag contains a subset of communities:

$$B_r^j = \{C \in \mathcal{C}; \lceil R_r(C) \cdot b/|\mathcal{C}| \rceil = j/|\mathcal{C}|\} \qquad \text{for } j = 1, 2, \ldots, B, \tag{19}$$

where $\lceil \cdot \rceil$ is the ceiling operator. It is possible that the last bag $B_r^B$ contains more than $|\mathcal{C}|/B$ communities. Within a bag, the ordering of communities is not important. Additionally, each community in bag $B_r^j$ has the same value of the importance weight $K_r^j$, which we explain next.

**Two-stage bootstrapping process.** The approach consists of two stages:

1. Compute the importance weights for each ranked list using the current aggregated prioritization,

2. Re-aggregate the ranked lists based on the importance weights computed in the previous stage.

After initializing the aggregated prioritization, the approach alternates between the two stages until no changes in the aggregated prioritization are observed.

**Stage 1: Estimating importance weights of bags.** In each iteration, the bootstrapping approach uses the current aggregated prioritization to re-compute the importance weight for each bag. This is done as follows. First, a temporary gold standard is constructed based on top ranked communities in the current aggregated prioritization. A temporary gold standard $T$ is a set containing $\pi|\mathcal{C}|$ communities that rank at the top in the current aggregated prioritization $R$:

$$T = \{C \in \mathcal{C}; R(C) \leq \pi|\mathcal{C}|\},$$

where $R(C)$ is the rank of community $C$ in the current aggregated prioritization $R$. Here, $\pi$, $0 < \pi < 1$, is a parameter representing the fraction of highest ranked communities used to construct the temporary gold standard $T$.

We formulate the importance weight for each bag following the Bayes factor formulation given in Supplementary Note 4.2. The importance weight $K_r^j$ for ranked list $R_r$ and bag $j$ compares the hypotheses $H_{r,j}^1$ and $H_{r,j}^2$ by evaluating the evidence in favor of hypothesis $H_{r,j}^1$. The hypotheses $H_{r,j}^1$ and $H_{r,j}^2$ are defined as in Supplementary Note 4.2 and have the following meaning:

$H_{r,j}^1$ – Bag $B_r^j$ matches the temporary gold standard $T$,

$H_{r,j}^2$ – Bag $B_r^j$ does not match the temporary gold standard $T$.

The importance weight $K_r^j$ is the ratio of the posterior odds of hypothesis $H_{r,j}^1$ to its prior odds:

$$K_r^j = \frac{p(H_{r,j}^1|R_r)}{p(H_{r,j}^2|R_r)} \frac{p(H_{r,j}^2)}{p(H_{r,j}^1)}. \tag{20}$$

Let us denote by $N_r^j$ the overlap between communities assigned to bag $B_r^j$ and communities in the temporary gold standard $T$, that is, $N_r^j = T \cap B_r^j$. This means that, in each iteration, $N_r^j$ contains all the temporarily gold standard communities that rank as the $j$-th bag in ranked list $R_r$. Following Eq. (20), we calculate the importance weight $K_r^j$ for ranked list $R_r$ and bag $j$ as:

$$K_r^j = \frac{|N_r^j| + 1}{|B_r^j| - |N_r^j| + 1} \cdot \frac{1-\pi}{\pi}. \tag{21}$$

Comparing the formula for the Bayes factor in Eq. (20) with the formula in Eq. (21), we can see the following. Equation (21) approximates the probability $p(H^1_{r,j}|R_r)$ by the fraction of communities in bag $B^j_r$ that are in the temporary gold standard $T$, $p(H^1_{r,j}|R_r) = (|N^j_r|+1)/(|B^j_r|+1)$. Similarly, Equation (21) approximates the probability $p(H^2_{r,j}|R_r)$ by the fraction of communities in bag $B^j_r$ that are not in the gold standard $T$, $p(H^2_{r,j}|R_r) = (|B^j_r|-|N^j_r|+1)/(|B^j_r|+1)$. Additionally, a smoothing value of one is added to prevent a division by zero.

It can also be seen from Eq. (21) that $\pi$, defined above as the relative size of temporary gold standard $T$, $\pi = |T|/|\mathcal{C}|$, actually corresponds to the prior probability of hypothesis $H^1_{r,j}$.

**Stage 2: Aggregating ranked lists.** In the second stage of the bootstrapping process, the approach aggregates the ranked lists based on the calculated importance weights. Following the rank aggregation model presented in Supplementary Note 4.3, the ranked lists are aggregated according to Eq. (17). More concretely, ranked lists are combined into the aggregated prioritization $R$ using the formula:

$$R(C) = \sum_r \log K_r^{i_r(C)} R_r(C), \tag{22}$$

where $K_r^{i_r(C)}$ is the importance weight of the bag $i_r(C)$ to which community $C$ is assigned in ranked list $R_r$. Here, $R(C)$ represents the aggregated rank of community $C$. Note that the aggregation formula uses the log importance weights, which correspond to predictive scores[26,28] that favor those bags in the ranked lists that better match the temporary gold standard.

Upon convergence of the two-stage bootstrapping procedure, the normalized value $R(C)$ gives the final aggregated rank of community $C$.

## Supplementary Note 5 CRANK approach

Following the presentation of the formal aspects of our approach for prioritizing network communities, we proceed by describing the complete CRANK algorithm.

### Supplementary Note 5.1 Overview of CRANK

The CRANK method consists of four steps. (1) First, a community detection algorithm is run on the network to identify communities. (2) In the second step, four community prioritization metrics are computed for each of the detected communities. This step yields four lists, each list containing scores of all communities for one prioritization metric. Scores in each list are then converted into ranks, producing ranked lists of communities. (3) In the third step, ranked list are aggregated resulting in the aggregated prioritization of communities. (4) Finally, in the fourth step, CRANK prioritizes the communities by ranking them by their decreasing aggregated score.

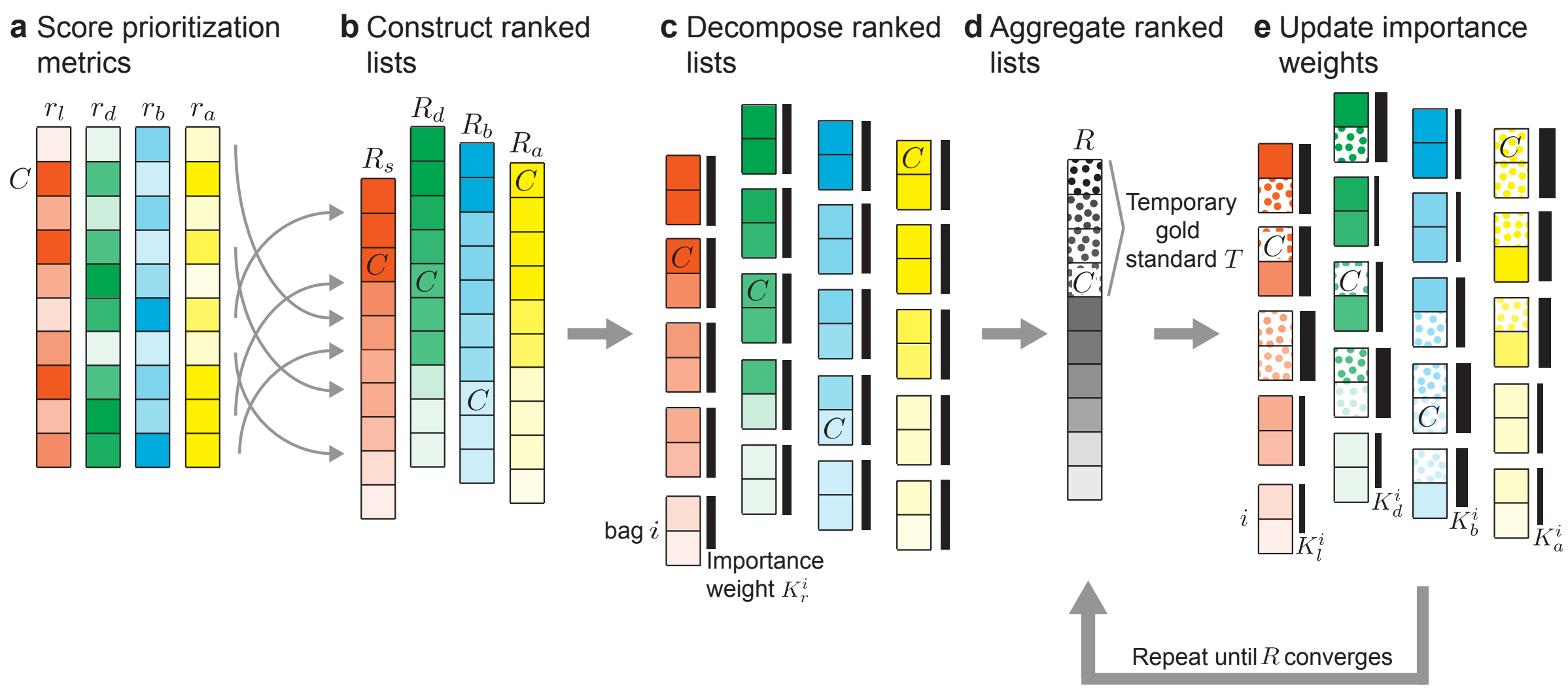

Supplementary Figure 1: **Community prioritization metrics and their combination.** Outlined are the second and the third step of CRANK (see Supplementary Note 5.1). **(a)** CRANK computes a score $r_f(C)$ for each detected community $C$ and each community prioritization metric $r_f$ (where $f$ is one of network structural features: "$l$" – likelihood, "$d$" – density, "$b$" – boundary, "$a$" – allegiance). **(b)** Scores $r$ are then sorted in a decreasing order into ranked list $R_r$. **(c)** Each ranked list is decomposed into equally sized bags. An importance weight $K_r^i$ (black vertical strip) is associated with each bag $i$ and each ranked list $R_r$. The weights are initially equal, denoting the aggregated prioritization $R$ as an equally weighted average of community ranks $R_l$, $R_d$, $R_b$ and $R_a$ at this point. **(d)** The highest ranked communities in $R$ form a temporary gold standard $T$ (dotted cells), which is used to update importance weights in the follow up steps. **(e)** For each bag $i$ and each ranked list $R_r$, a new importance weight $K_r^i$ is calculated according to the current aggregated prioritization $R$ using Eq. (21). CRANK proceeds by updating the aggregated prioritization $R$ according to the revised importance weights using Eq. (22). Calculations in **(d)** and **(e)** are repeated until aggregated prioritization $R$ converges.

We proceed by explaining the aggregation phase (*i.e.*, the third step) in more detail (Supplementary Figure 1). At the start, CRANK sorts scores from each metric $r$ (Supplementary Figure 1a) into a list $R_r$ of community ranks (Supplementary Figure 1b), and it then partitions these lists into bags, which are equally sized sets of communities described in Supplementary Note 4.4.2 (Supplementary Figure 1c). Next, an initial aggregated prioritization of communities is generated as an equally weighted average of community ranks $R_l$, $R_d$, $R_b$ and $R_a$ (Supplementary Figure 1d). The algorithm then iterates until the aggregated prioritization converges (*i.e.*, community ranks do not change between two consecutive iterations) or the maximum number of iterations is reached (Supplementary Figure 1d-e).

In each iteration, a set of the highest ranked communities (*i.e.*, a "temporary gold standard"

described in Supplementary Note 4.4.2) is formed based on the current aggregated prioritization $R$ (Supplementary Figure 1d). The approach then calculates the importance weight $K_r^i$ for each bag $i$ and each ranked list $R_r$ using Eq. (21) by considering communities in the temporary gold standard as a point of reference (Supplementary Figure 1e). CRANK determines the importance weight of a bag based on the number of communities in the bag that are also contained in the temporary gold standard. CRANK applies Tukey's running median smoothing procedure[33] to make the importance weights robust. Finally, CRANK uses Eq. (22) to update the current aggregated prioritization $R$. This is done by combining community ranks $R_l$, $R_d$, $R_b$ and $R_a$ into the aggregated prioritization $R$ according to the importance weights. Repeating this procedure to iteratively refine the aggregated prioritization $R$ underlies CRANK.

### Supplementary Note 5.2 CRANK algorithm

A complete description of the CRANK algorithm follows.

- **Input:** Network $G(\mathcal{V}, \mathcal{E})$, community detection algorithm $A$
- **Parameters:** Network perturbation intensity $\alpha$, number of bags $B$, relative size of temporary gold standard $\pi$
- **Output:** Aggregated prioritization $R$

1. **Step: Community detection**
   - Apply community detection algorithm $A$ on network $G$ to detect communities $\mathcal{C}$:

$$M(\mathcal{V}, \mathcal{E}, \mathcal{C}) \leftarrow A(G)$$

2. **Step: Community prioritization metrics**
   - Compute edge probabilities under $\alpha$-perturbation of network $G$ using $M(\mathcal{V}, \mathcal{E}, \mathcal{C})$ (Eqs. (1, 2, 3, 4))
   - For each detected community $C \in \mathcal{C}$, compute the scores:
     - likelihood $r_l(C; \alpha)$
     - density $r_d(C; \alpha)$
     - boundary $r_b(C; \alpha)$
     - allegiance $r_a(C; \alpha)$
   - For each metric $r \in \{r_l, r_d, r_b, r_a\}$, form a ranked list $R_r$ such that $R_r(C)$ is the rank (*i.e.*, position) of community $C$ in $R_r$:

$$R_r(C) = 1 + \sum_{D \neq C} I(r(C; \alpha) \leq r(D; \alpha)) \tag{23}$$

3. **Step: Combining community prioritization metrics**

- Decompose each ranked list $R_r$ into $B$ equally sized bags such that $j$-th bag contains a set of communities:

$$B_r^j = \{C \in \mathcal{C}; \lceil R_r(C) \cdot B/|\mathcal{C}| \rceil = j/|\mathcal{C}|\} \qquad \text{for } j = 1, 2, \ldots, B$$

- Initialize the importance weights $K_r^j$ for each ranked list $R_r$ and each bag $j$ as $K_r^j = \frac{1}{4}$
- Repeat until the aggregated prioritization $R$ does not change between two consecutive iterations or a maximum number of iterations is reached:
  - Construct the aggregated prioritization $R$ by combining the ranked lists as:

$$R(C) = \sum_r \log K_r^{i_r(C)} R_r(C), \tag{24}$$

    where $K_r^{i_r(C)}$ is the importance weight of the bag $i_r(C)$ to which community $C$ is assigned in ranked list $R_r$
  - Convert the aggregated prioritization $R$ into rank order as:

$$R(C) \leftarrow 1 + \sum_{D \neq C} I(R(C) \leq R(D)) \tag{25}$$

    To deal with ties, the average of the ranks that would have been assigned to all the tied communities is assigned to each community
  - Form a temporary gold standard $T$ consisting of $\pi|\mathcal{C}|$ highest ranked communities in $R$:

$$T = \{C \in \mathcal{C}; R(C) \leq \pi|\mathcal{C}|\},$$

    where $R(C)$ is the rank of community $C$ in the aggregated prioritization $R$
  - Update the importance weight $K_r^j$ for each metric $r$ and each bag $j$ using the formula:

$$K_r^j = \frac{|N_r^j| + 1}{|B_r^j| - |N_r^j| + 1} \cdot \frac{1 - \pi}{\pi} \tag{26}$$

    where $N_r^j = T \cap B_r^j$
  - Smooth the importance weights of each ranked list $R_r$ and bag $B_r^j$ using the Tukey's running median procedure[33] with window size three:

$$K_r^j = \begin{cases} \text{median}(K_r^{j-1}, K_r^j, K_r^{j+1}) \text{ if } j \neq 1, B \\ \text{median}(K_r^1, K_r^2, 3K_r^2 - 2K_r^3) \text{ if } j = 1 \\ \text{median}(K_r^b, K_r^{b-1}, 3K_r^{b-1} - 2K_r^{b-2}) \text{ if } j = B \end{cases}$$

    Continue to apply the median smoothing to the importance weights of metric $r$ until no more changes are observed

4. **Step: Generating community ranking**

- Return the rank-ordered aggregated prioritization $R$

**Community detection algorithms.** CRANK can be applied to communities detected with a number of statistical community detection methods. Examples include community detection methods based on Affiliation Graph Model[8,9], Stochastic Block Model[7], Latent Feature Graph Model[10–16,34] and Attributed Graph Model[17–19]. Additionally, CRANK works with non-statistical methods like modularity optimization and spectral methods, where edge probabilities are given by an auxiliary network model.

**Other parameters of** CRANK**.** The CRANK algorithm has three parameters: network perturbation intensity $\alpha$, number of bags $B$, and relative size of temporary gold standard $\pi$.

We find empirically that CRANK rank aggregation method always converges in less than 20 iterations of the algorithm and it takes on average less than 10 iterations for the aggregated ranking to converge. In the algorithm we track the change of the aggregated ranking between two consecutive iterations and stop the algorithm when no change in the ranking is observed. At that point the ranking has completely stabilized, it will not change in future iterations, and thus the aggregation is said to converge. Although we use the most strict stopping criterion in our experiments, we note that we have not observed any convergence issues, even when aggregating large ranked lists with more than ten thousand communities.

We find that for rank-based aggregation of CRANK metrics, a choice for bag size of around 50 is appropriate. That means that the number of bags is set to $B = |\mathcal{C}|/50$, where $\mathcal{C}$ denotes the number of communities detected in a network, and that all bags are of equal size. We use that value for the number of bags in all experiments reported in the paper, unless the experiment involves prioritizing fewer than 50 communities. In the latter case, we require at least three bins.

We have evaluated the sensitivity of CRANK to different perturbation intensities $\alpha$ of a network. All results reported in this paper are obtained by assuming a small perturbation of the network structure, $\alpha = 0.15$. This means that CRANK metrics capture the magnitude and the robustness of network structural features, which is important for good prioritization performance.

We have investigated a number of values for the relative size $\pi$ of temporary gold standard. We observe that setting the relative size to $\pi = 0.05$ performs well across different datasets and community detection algorithms and we use that value in all our experiments.

### Supplementary Note 5.3 Computational time complexity of CRANK

We separately analyze computational complexity of each of the fours CRANK steps.

The runtime of the first step is the time needed to detect communities in the network $G$. We denote this time as $\mathcal{O}(A)$ . In the second step, CRANK calculates community prioritization metrics for all detected communities. This step takes time $\mathcal{O}(|\mathcal{C}| \cdot |\mathcal{E}| + |\mathcal{C}| \cdot \max_i |C_i|^2 + |\mathcal{C}| \cdot |\mathcal{E}|)$, where the first term is due to computation of edge probabilities based on CRANK network perturbation model, the second term is due to computation of likelihood metric scores and the third term is due to computations of community density, boundary, and allegiance metric scores. This means that computing metric scores in the second step requires time, which is linear in the size of network $G$. The third step is computationally straightforward and requires $\mathcal{O}(n_{\text{iter}} \cdot |\mathcal{C}| \cdot \log |\mathcal{C}|)$ time, where $n_{\text{iter}}$ is the maximum number of iterations needed for aggregation of community metric scores. We note that $n_{\text{iter}} < 20$ was sufficient for convergence of the aggregated prioritization in all our studies. The fourth step of CRANK is similarly straightforward: it involves sorting the communities based on their overall score in the aggregated prioritization $R$. Altogether, the time complexity of CRANK is the sum of the times needed to complete all four steps of CRANK algorithm, $\mathcal{O}(A + |\mathcal{C}| \cdot |\mathcal{E}| + |\mathcal{C}| \cdot \max_i |C_i|^2 + n_{\text{iter}} \cdot |\mathcal{C}| \cdot \log |\mathcal{C}|)$.

Notice that, in the second step, a traditional, explicit approach to computing edge probabilities in the perturbed network would first perform many physical perturbations of the input network and would then run a community detection algorithm on each of the perturbed networks. This procedure would take time $\mathcal{O}(n_{\text{pert}} \cdot (|\mathcal{E}| + \mathcal{O}(A)))$, where $n_{\text{pert}}$ counts network perturbations, typically[2] $n_{\text{pert}} \gg 100$. However, by computing edge probabilities in the perturbed network analytically rather than empirically, CRANK only needs time $\mathcal{O}(|\mathcal{C}| \cdot |\mathcal{E}|)$, which substantially increases CRANK's scalability for large networks.

CRANK naturally allows for parallelization, which further increases its scalability. Calculations of prioritization metric scores are independent for each metric and each community and thus can be computed in parallel.

## Supplementary Note 6 Details about datasets

Next, we describe datasets considered in this study.

### Supplementary Note 6.1 Network datasets

We consider twelve networks from biological, social, and information realms (Supplementary Table 1).

Supplementary Table 1: **Statistics of network datasets.** $N$: number of nodes, $E$: number of edges, $C$: number of ground-truth communities, $S$: average ground-truth community size, $A$: average ground-truth community memberships per node. The Medical drugs network has three types of ground-truth information: Text[35], Chemistry[35], Epistasis[36], given in the form of drug-drug relationships ($C$: number of drug-drug relationships, $A$: average relationships per node). The HSDN network has three types of ground-truth information: Pathway[37], Genes[37], Chemistry[37], also given in the form of disease-disease relationships ($C$: number of disease-disease relationships, $A$: average relationships per node).

| Dataset | $N$ | $E$ | $C$ | $S$ | $A$ |
|---|---|---|---|---|---|
| Google+[38] | 0.11 m | 27.3 m | 437 | 143.5 | 0.3 |
| Facebook[38] | 4.04 k | 0.18 m | 193 | 28.8 | 1.4 |
| Twitter[38] | 81.31 k | 3.5 m | 3.1 k | 15.5 | 0.4 |
| DBLP[38] | 0.32 m | 1.0 m | 13.5 k | 429.8 | 2.6 |
| Amazon[38] | 0.33 m | 0.9 m | 151 k | 99.9 | 14.8 |
| Human Net[39] | 16.24 k | 0.48 m | 4.1 k | 16.9 | 8.4 |
| Human IntAct[40] | 23.68 k | 0.11 m | 4.3 k | 14.7 | 9.0 |
| Yeast GI[41] | 4.46 k | 0.17 m | 1.2 k | 9.8 | 8.9 |
| Human BioGRID[42] | 19.56 k | 0.17 m | 3.9 k | 15.9 | 9.5 |
| Human STRING[43] | 5.42 k | 50.8 k | 815 | 59.9 | 9.0 |
| | | | 28.6 k | / | 43.5 |
| Medical drugs[36] | 1.32 k | 16.7 k | 1.4 k | / | 2.1 |
| | | | 21.8 k | / | 33.1 |
| | | | 164.9 k | / | 268.8 |
| HSDN[44] | 1.23 k | 6.5 k | 172.5 k | / | 281.1 |
| | | | 145.1 k | / | 236.6 |

We consider five gene networks: Human Net, Human Int Act, Yeast GI, Human BioGRID, and Human STRING. Human STRING is a protein-protein interaction network of experimentally determined interactions between human proteins from STRING v9.1 database[43]. The nodes are limited to proteins associated with biological pathways in the Reactome database[45]. Human Net[39] is a human-specific gene network that combines gene co-citation, gene co-expression, curated physical protein interactions, genetic interactions, and co-occurrence of protein domains from four species. We also consider human genetic and physical interaction network (Human IntAct)[40], experimentally derived genetic interaction network in yeast *S. cerevisiae* (Yeast GI)[41], and human protein-protein interaction network (Human BioGRID)[42]. All networks are provided as part of relevant publications.

We consider two non-gene/non-protein networks: Medical drugs and HSDN. Medical drug network contains drugs approved by the U.S. Food and Drug Administration (FDA), which are listed in the DrugBank database[36]. Two drugs are linked in the network if they have at least one target protein in common. The HSDN network contains human diseases, where two diseases are linked if their clinical symptoms are significantly similar[44]. Both networks are provided as part of relevant publications.

We consider five social and information networks representing standard benchmark datasets in network science. We consider a collaboration network from the DBLP computer science bibliography[46], the Amazon product co-purchasing network[46], and a collection of ego-networks from online social networks of Google+, Twitter and Facebook[13]. All networks are downloaded from the SNAP database[38] and are publicly available at: `http://snap.stanford.edu/data`.

### Supplementary Note 6.2 Ground-truth community datasets

For all considered networks, we have explicit *ground-truth* membership of nodes to communities (Supplementary Table 1). This means that in all networks ground-truth community memberships of nodes have been externally validated and verified.

Ground-truth communities for the Human STRING protein-protein interaction network is given by curated biological pathways from the Reactome database[45]. For other gene/protein networks we obtain ground-truth communities from the Gene Ontology[47] in the form of gene groups that correspond to biological processes, cellular components and molecular functions (see Supplementary Note 7.4). For the HSDN network we have three types of ground-truth information from the Comparative Toxicogenomics Database: information about molecular pathways that are common to disease pairs[37], knowledge about disease genes that are common to disease pairs[37], and chemical associations that are common to disease pairs[37]. For the medical drug network we also obtain three types of ground-truth information: text associations between chemicals from the STITCH database[35], drug-drug relationships from the STITCH[35] based on similarity of drug's chemical structure, and drug-drug interactions from the DrugBank[36].

In the Amazon product co-purchasing network, ground-truth communities are defined by product categories on the Amazon website[13,46]. In the DBLP network, ground-truth communities are defined by publication venue,*e.g.*, journal or conference, meaning that authors who published to a certain journal or conference form a ground-truth community[46]. In the online social networks, ground-truth communities are defined by users' social circles[13].

## Supplementary Note 7 Details about experimental setup

We describe community detection approaches that are used to find network communities, which we then prioritize. We then describe the experimental design and the metrics used for performance evaluation.

### Supplementary Note 7.1 Community detection methods

We use the following community detection methods: CoDA (Communities through Directed Affiliations)[9], BigCLAM[8] and MMSB (Mixed Membership Stochastic Blockmodels)[16,34]. These methods implement different statistical models of community detection and are hence appropriate for use with CRANK. We use publicly available implementations of the methods. Implementations of CoDA and BigCLAM are provided as part of the SNAP library[48]. MMSB is implemented in Chang *et al.*[49] Values for model parameters of the methods were selected based on method's authors recommendation. Estimates of edge and node-community membership probabilities, which are needed by CRANK, were obtained with tools for examining posterior distributions, which are included in the SNAP[48] and in Chang *et al.*[49].

### Supplementary Note 7.2 Baseline community metrics

For comparison we consider Conductance[50] and Modularity[51] community scoring functions. In order to make a higher value better, we reverse Conductance as $(1-\text{Conductance})$. We also consider two simple baselines: random prioritization of communities, and prioritization by the increasing size of communities.

Conductance of a community $C$ is defined as $\text{Conductance}(C) = |B_C|/(2|E_C| + |B_C|)$, where $E_C$ are edges within community $C$, $E_C = \{(u,v) \in \mathcal{E}; u \in C, v \in C\}$, and $B_C$ are edges leaving community $C$ (*i.e.*, community's edge boundary), $B_C = \{(u,v) \in \mathcal{E}; u \in C, v \notin C\}$. If Conductance is used for community ranking then the highest ranked communities are those with the smallest fraction of total edge volume pointing outside them. Modularity of a community $C$ is defined as $\text{Modularity}(C) = 1/4(|E_C| - E[|E_C|])$. It measures the difference between the number of edges in a community and the expected number of such edges in a random graph with identical degree distribution. In prioritization, modularity prefers communities that are denser (*i.e.*, with many internal edges) than what is expected under the configuration random network model[2,51].

We also considered several other community scoring functions[46]: Flake-ODF (Out-Degree

Fraction), Cut Ratio, TPR (Triangle Participation Ratio) and FOMD (Fraction Over Median Degree). We reversed metrics Cut Ratio and Flake-ODF as $(1 - \text{Cut Ratio})$ and $(1 - \text{Flake-ODF})$, respectively, to make a higher value indicate higher priority. In our experiments these scoring functions were typically outperformed by either Conductance or Modularity or both. When this was the case, their results are not reported.

We also evaluate CRANK against its several simplified variants:

- We compare different subsets of CRANK's prioritization metrics to each other. For example, we use CRANK's rank aggregation method to aggregate the scores of community likelihood, community density and community boundary, but we leave out the scores of community allegiance.
- We compare CRANK's rank aggregation method to methods that aggregate metric scores via simple quadratic mean, Borda method[52], Footrule approach[53], and Pick-a-Perm[54].
- We compare CRANK's prioritization metrics to different combinations of the baseline scoring functions. For example, we use CRANK's rank aggregation method to combine Cut Ratio, Conductance, TPR and FOMD scoring functions.

### Supplementary Note 7.3 Prioritization performance evaluation

Next we describe gold standard rankings that are used to evaluate prioritization performance.

The intuition behind our experiments is the following: We want communities with higher prioritization scores (*i.e.*, communities that rank closer to the top of a ranked list) to provide a more accurate reconstruction of the ground-truth communities. More precisely, given only a network, we first detect communities and then prioritize them with the goal to establish which detected communities are the most accurate without actually knowing the ground-truth community labels. A perfect prioritization would order communities by decreasing accuracy, such that detected communities, which best match the ground-truth communities, are ranked at the top.

We would like to note that community detection methods detect communities using only network structure and community prioritization methods prioritize communities using only information about community structure. This means that community detection and prioritization methods do not consider any external metadata or labels. We can thus quantitatively evaluate performance of a prioritization method by comparing community rankings generated by the method with gold standard community rankings determined by the ground-truth information.

We evaluate prioritization of communities by quantifying its correspondence with the *gold standard ranking of communities*. We determine the gold standard ranking by computing the accuracy of every detected community by matching it to ground-truth communities. We adopt the following evaluation procedure previously used in Yang *et al.*[46]: Every detected community $C$ is matched with its most similar ground-truth community $C^*$. Given $\mathcal{C}^*$, the set of all ground-truth communities that is explicitly provided by an external data resource, such as the SNAP[38], $C^*$ is defined as: $C^* = \arg\max_{D^* \in \mathcal{C}^*} \delta(D^*, C)$, where $\delta(D^*, C)$ measures the Jaccard similarity between ground-truth community $D^*$ and detected community $C$. $C^*$ is thus the ground-truth community that is the best match for a given detected community $C$. The accuracy of community $C$ is simply the Jaccard similarity, $\delta(C^*, C)$, between $C$ and its corresponding ground-truth community $C^*$. The gold standard ranking is then defined as the ranking of the detected communities by decreasing accuracy.

A perfect prioritization matches the gold standard ranking exactly and ranks communities in decreasing order of accuracy. In this case, the Spearman's rank correlation $\rho$ between the gold standard ranking and the estimated prioritization is one. The Spearman's rank correlation $\rho$ is close to zero when the prioritization of communities does not carry any signal, and negative when the predicted prioritization tends to order the detected communities by the increasing accuracy rather than by the decreasing accuracy.

### Supplementary Note 7.4 Functional enrichment analysis

Functional enrichment analysis[55] is an established computational procedure in biology for the rigorous assessment of statistical significance of gene sets. The input to functional enrichment analysis consists of (1) a gene set, *i.e.*, a community detected in a gene network given by its member genes, and (2) known gene functional annotation data. The output is statistical significance of their association.

We obtain known sets of functionally related genes from the Gene Ontology (GO)[47]. GO terms are organized hierarchically such that higher level terms, *e.g.*, "regulation of biological process", are assigned to more genes and more specific descendant terms, *e.g.* "positive regulation of eye development", are related to parent by "is_a" or "part_of" relationships. We consider high confidence experimentally validated GO annotations (*i.e.*, annotations associated with the evidence codes: EXP, IDA, IMP, IGI, IEP, ISS, ISA, ISM or ISO) that cover all three aspects in the GO: biological processes, molecular functions and cellular components. Since the obtained GO data only contain the most specific annotations explicitly, we retrieve the relevant GO annotations and propagate them upwards through the GO hierarchy, *i.e.*, any gene annotated to a certain GO term is also explicitly included in all parental terms[56,57].

We evaluate the significance of the association between each detected community and the GO

using PANTHER tool[58] in February 2015 (*i.e.*, "PANTHER Over-representation Analysis" using Fisher's exact test). The Bonferroni correction was used to account for multiple testing. Given a detected community, the over-representation analysis tests which GO terms are most associated with the community and evaluates if their association is significantly different (p-value $< 0.05$, Bonferroni correction for multiple testing) from what is expected by chance. The basic question answered by this test is: when sampling $X$ genes (a detected community) out of $N$ genes (all nodes in the network), what is the probability that $x$ or more of these genes belong to a particular GO term shared by $n$ of the $N$ genes in the network? The Fisher's exact test answers this question in the form of a p-value. We say that a community is *functionally enriched in a given GO term* if it is significantly associated with that GO term. Intuitively, this means that a community contains surprisingly large number of genes that perform the same cellular function, are located in the same cellular component, or act together in the same biological process, as defined by a given GO term. We say that a community is *functionally enriched* if it is functionally enriched in at least $k$ GO terms, where $k$ is pre-selected value (*i.e.*, $k_C = |C|$ for a community $C$).

To evaluate the quality of community prioritization we report how many communities that rank among the top 5% of all communities are functionally enriched. A larger number of functionally enriched communities at the top of a community ranking indicates better prioritization performance.

## Supplementary Note 8 Experiments on CRANK and its properties

In this note, we investigate CRANK's properties. We study CRANK metrics and CRANK rank aggregation method, two major components of CRANK approach. We start by applying CRANK in conjunction with different community detection methods (Supplementary Note 8.1) and evaluating CRANK's sensitivity to network perturbation intensity (Supplementary Note 8.2). We then evaluate the contribution of each CRANK metric towards the performance of CRANK (Supplementary Note 8.3). We then compare CRANK against combinations of baseline community metrics in order to better understand the impact of CRANK metrics on performance (Supplementary Note 8.4). Finally, we evaluate how the proposed rank aggregation method performs in comparison to existing rank aggregation methods (Supplementary Note 8.5).

All experiments reported in this note are done on social and information networks (Supplementary Note 6.1) because of available high-quality (*i.e.*, complete) ground-truth information that is used to evaluate prioritization performance.

## Supplementary Note 8.1 CRank in conjunction with different community detection methods

We consider five social and information networks. For each network, we used a community detection method to detect communities, and then we prioritized the detected communities using CRank. To demonstrate that CRank can be used with any community detection method, we here use CRank in conjunction with three state-of-the-art community detection methods (*i.e.*, CoDA[9], BigCLAM[8], and MMSB[16], see Supplementary Note 7.1).

For the purpose of evaluation we consider networks with ground-truth information on communities[46]. Notice that this information is not available to methods during community detection or community prioritization. However, it enables us to compile a gold standard ranking of communities, which ranks communities based on how well they reconstruct ground-truth, *i.e.*, externally validated, communities. Spearman's rank correlation $\rho$ is used to measure how well a generated ranking approximates the gold standard ranking (see Supplementary Note 7.3). We compare performance of CRank to alternative metrics potentially useful for prioritization: modularity, conductance, and random prioritization.

Tables Supplementary Table 2-Supplementary Table 4 show the performance of CRank and other baseline community metrics on five networks under the BigCLAM, MMSB, and CoDA community detection methods. Overall, we find that across all datasets and community detection methods, CRank is always the best performing method to prioritize communities. CRank outperforms Modularity by up to 128% and generates on average 57% more accurate community rankings as measured by the Spearman's rank correlation between a generated ranking and the gold standard ranking. Similarly, CRank outperforms Conductance by up to 107% and generates on average 38% more accurate community rankings. Furthermore, CRank performs on average 32% better than the second best community metric. The second best performing community metric changes considerably across the datasets, while CRank always performs best, suggesting that it can effectively exploit the network structural features to become aware of a particular configuration of a dataset and a community detection model. CRank outperforms other community metrics, and we hypothesize that the scoring functions of those metrics are unable to model heterogeneity of datasets and community detection algorithms.

Importantly, results in Tables Supplementary Table 2-Supplementary Table 4 show that CRank performs substantially better than the approach, which is nowadays typically employed when no other domain-specific meta or label information besides the network structure is available (*i.e.*, Random). On average, CRank improves the random ordering of the detected communities by more than 10 folds as measured by the Spearman's rank correlation. These results suggest that the notion of community priority employed by CRANK agrees well with a gold standard ranking that is measured via ground-truth community information; in fact, CRANK does so in a completely unsupervised manner.

Supplementary Table 2: **Prioritization of communities detected by the BigCLAM method.** We measure Spearman's rank correlation between the generated ranking of communities and the gold standard ranking of communities. Higher values indicate better performance. Communities were detected by the BigCLAM algorithm[8] and prioritized using one of four different approaches. Higher value indicates better performance.

| Method | Facebook | Amazon | Google+ | Twitter | DBLP |
|---|---|---|---|---|---|
| Random | 0.003 | -0.019 | 0.089 | 0.031 | 0.025 |
| Modularity | 0.225 | 0.158 | 0.307 | 0.277 | 0.252 |
| Conductance | 0.287 | 0.216 | 0.293 | 0.319 | 0.333 |
| CRANK | 0.342 | 0.278 | 0.325 | 0.442 | 0.358 |

Supplementary Table 3: **Prioritization of communities detected by the MMSB method.** We measure Spearman's rank correlation between the generated ranking of communities and the gold standard ranking of communities. Higher values indicate better performance. Communities were detected by the MMSB algorithm[16] and prioritized using one of four different approaches. Higher value indicates better performance.

| Method | Facebook | Amazon | Google+ | Twitter | DBLP |
|---|---|---|---|---|---|
| Random | 0.087 | 0.013 | -0.016 | -0.010 | 0.008 |
| Modularity | 0.281 | 0.183 | 0.295 | 0.302 | 0.251 |
| Conductance | 0.329 | 0.218 | 0.329 | 0.417 | 0.281 |
| CRANK | 0.356 | 0.295 | 0.384 | 0.439 | 0.371 |

Based on these results we conclude that CRANK consistently achieves good performance measured in terms of Spearman's rank correlation on the ground-truth community information. Furthermore, the results indicate that CRANK can be successfully applied to popular and state-of-the-art community detection methods.

Supplementary Table 4: **Prioritization of communities detected by the CoDA method.** We measure Spearman's rank correlation between the generated ranking of communities and the gold standard ranking of communities. Higher values indicate better performance. Communities were detected by the CoDA algorithm[9] and prioritized using one of four different approaches. Higher value indicates better performance.

| Method | Facebook | Amazon | Google+ | Twitter | DBLP |
|---|---|---|---|---|---|
| Random | -0.016 | 0.012 | 0.005 | -0.013 | 0.009 |
| Modularity | 0.149 | 0.195 | 0.166 | 0.213 | 0.301 |
| Conductance | 0.257 | 0.161 | 0.187 | 0.226 | 0.212 |
| CRANK | 0.340 | 0.267 | 0.312 | 0.465 | 0.411 |

### Supplementary Note 8.2 Network perturbation intensity in CRANK

We evaluate the sensitivity of CRANK to different perturbation intensities $\alpha$ of a network. Recall that CRANK defines prioritization metrics as follows. Given a structural feature $f$, CRANK defines prioritization metric $r_f$ such that it captures the magnitude of feature $f$ in the network as well as the change in the value of $f$ between the network and its $\alpha$-perturbed version: $r_f(C;\alpha) = f(C)/(1 + d_f(C, \alpha))$ (Supplementary Note 3). It can thus be expected that perturbation intensity $\alpha$ might influence CRANK's prioritization performance. We here vary the value for $\alpha$ and study its impact on CRANK's performance.

Supplementary Figure 2 shows the performance achieved on the Amazon, DBLP, and STRING networks for different values of perturbation intensity $\alpha$ varying from $\alpha = 0.05$ to $\alpha = 0.95$ with an increasing step of $0.05$. We observe that varying $\alpha$ influences the overall performance across different networks and community detection algorithms.

Results in Supplementary Figure 2 are consistent with the accepted definition of stability and robustness of community structure in networks[4,59–62]. It is generally posited[2,5] that, at the network level, significant community structure should be robust to *small perturbations of the network* (*i.e.*, for low values of $\alpha$). This notion corresponds to the robustness of community structure against noise and data incompleteness[6]. In other words, if a small change in the network can completely change the outcome of community detection algorithm, then the communities found should not be considered significant.

However, when perturbation intensity is beyond a certain threshold, *i.e.*, when the network is perturbed to such extent that it resembles a random network (*i.e.*, for large values of $\alpha$), then a good metric should assign community structure detected in the perturbed network a low score even if community structure in the original network is significant[2]. This notion corresponds to the specificity of the community structure that should be captured by a good community metric. Therefore, in CRANK, the adjusted prioritization metrics should be most informative for small values of perturbation intensity $\alpha$. This indeed holds for CRANK, as can be seen in Supplementary Figure 2. For values of perturbation intensity beyond a reasonably small threshold (*e.g.*, $\alpha = 0.3$), prioritization performance typically slowly deteriorates.

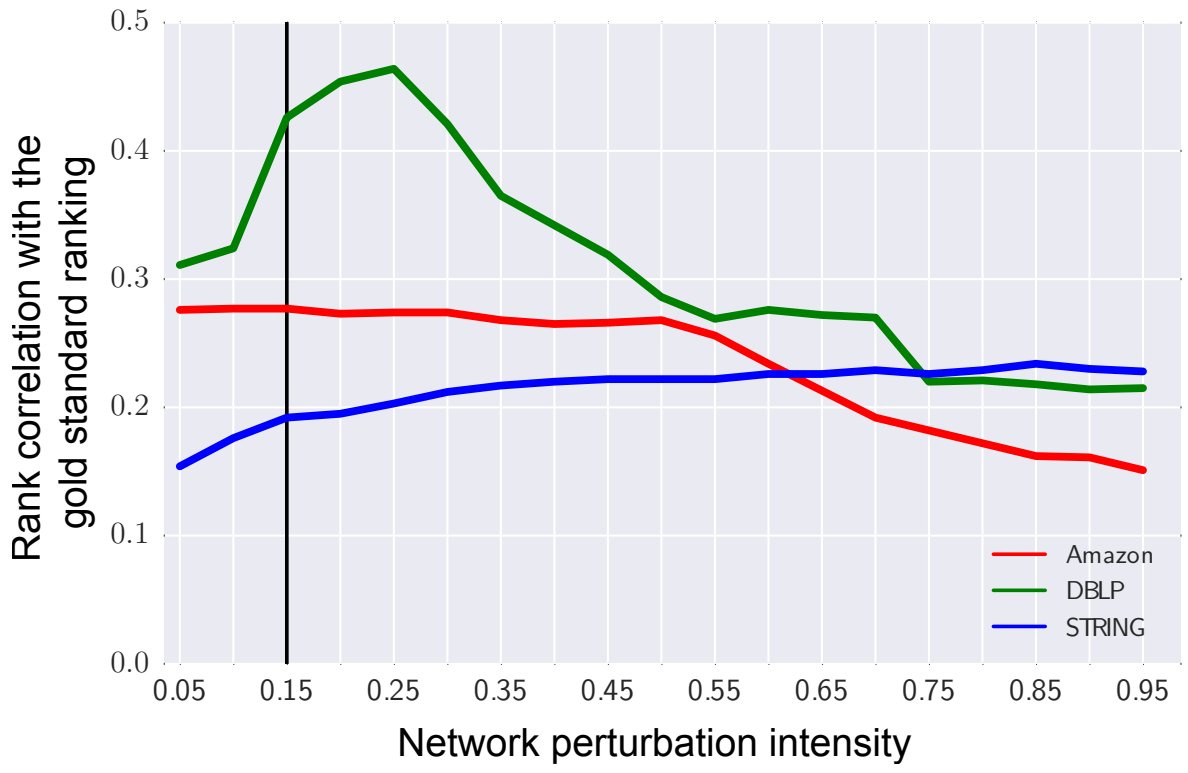

Supplementary Figure 2: **Prioritization performance of CRank when varying the value of network perturbation intensity.** Network communities are found using the CoDA community detection algorithm[9]. Perturbation intensity $\alpha$ affects the estimates of edge probabilities that are computed by CRank and used to analytically determine community structure in the perturbed network based on communities detected in the original network. All results of CRank reported in this paper are obtained by assuming a small perturbation of the network structure[2], $\alpha = 0.15$ (black vertical line).

An especially interesting case is to investigate CRANK's performance when $\alpha = 0$. When $\alpha = 0$, the formula for prioritization metric $r_f$ becomes $r_f(C; 0) = f(C)$. In other words, when $\alpha = 0$, prioritization metric $r_f$ only captures the magnitude of feature $f$ in the network. This means that the metric $r_f$ ignores any information, which comes from the change in the value of $f$ between the original network and its perturbed version.

On average, on Amazon, DBLP, and STRING networks we observe that setting $\alpha$ to $\alpha = 0$ results in 27% lower Spearman's rank correlation with the gold standard ranking as compared to Spearman's rank correlation when $\alpha = 0.15$. These findings suggest that both the magnitude and the robustness of network structural features have an important role in CRANK's performance.

Supplementary Table 5: **Complementarity of CRANK community prioritization metrics.** We leave out one of the four CRANK community prioritization metrics at a time and evaluate performance of the reduced CRANK by measuring Spearman's rank correlation between predicted prioritization of communities and the gold standard ranking of communities. We also report the accuracy of individual community metrics. Communities were detected by the CoDA algorithm[9]. The "\" sign denotes prioritization metric that was left out in the experiment. Higher value indicates better performance.

| Ranking method | Facebook | Amazon | Google+ | Twitter | DBLP | Average |
|---|---|---|---|---|---|---|
| $r_l$ | 0.277 | 0.187 | 0.247 | 0.211 | 0.309 | 0.246 |
| $r_d$ | 0.301 | 0.209 | 0.199 | 0.255 | 0.276 | 0.248 |
| $r_b$ | 0.245 | 0.178 | 0.188 | 0.288 | 0.213 | 0.222 |
| $r_a$ | 0.289 | 0.189 | 0.245 | 0.188 | 0.311 | 0.244 |
| CRANK $\setminus r_l$ | 0.303 | 0.203 | 0.262 | 0.423 | 0.312 | 0.300 |
| CRANK $\setminus r_d$ | 0.267 | 0.189 | 0.241 | 0.367 | 0.353 | 0.274 |
| CRANK $\setminus r_b$ | 0.278 | 0.193 | 0.283 | 0.316 | 0.347 | 0.283 |
| CRANK $\setminus r_a$ | 0.311 | 0.245 | 0.271 | 0.474 | 0.323 | 0.283 |
| CRANK | 0.340 | 0.267 | 0.312 | 0.465 | 0.411 | 0.359 |

In other words, high priority communities have high values of community prioritization metrics, which are also stable with respect to small perturbations of the network structure.

### Supplementary Note 8.3 Incremental contribution of CRANK metrics

We examine the degree of contribution of each of the four CRANK metrics to the final performance of CRANK. Recall that CRANK metrics are: (1) Community likelihood $r_l$, which scores each community based on the overall likelihood of edges and non-edges in the community; (2) Community density $r_d$, which scores each community based on the probability of community's internal network connectivity; (3) Community boundary $r_b$, which scores each community based on the sharpness of its edge boundary; and (4) Community allegiance $r_a$, which scores each community based on the difference between internal and external network connectivity of each community member.

We want to test whether the four CRANK metrics are truly necessary or would CRANK perform just as well with only a subset of them. To answer this question, we consider in turn different subsets of CRANK metrics and apply CRANK with each of the subsets.

Supplementary Table 5 shows that considering all CRANK metrics improved average Spear-man's rank correlation obtained by considering only one metric by 50%. It improved Spearman's rank correlation of the best single CRank metric considered in isolation by 45%. Additionally, all four CRank metrics performed on average 26% better than any subset of three metrics. These observations suggest that each prioritization metric by itself carries a substantial predictive signal, and that combining all the metrics results in superior performance. We hence conclude that the proposed metrics are complementary, and that good performance of CRank depends on consideration of all of them.

### Supplementary Note 8.4 Combinations of baseline community metrics

To better understand the impact of CRank aggregation method and CRank metrics on performance, we compare CRank against standard and commonly used community metrics. We evaluate the accuracy of community rankings obtained by combining six baseline community metrics as well as all combinations of five out of the six the metrics (*i.e.*, Cut Ratio[46], Conductance[50], TPR[46], FOMD[46], Flake-ODF[46], Modularity[51]; see Supplementary Note 7.2). Baseline community metrics in each combination are aggregated by averaging the metrics' scores.

Results are reported in Supplementary Table 6. We can learn two things by examining results of this experiment. First, comparing performance of the aggregated metric scores in Supplementary Table 6 with performance of the non-aggregated metric scores reveals that the aggregated metric scores consistently performed better than any one metric by itself. For example, aggregation of Conductance with FOMD, TPR, Cut Ratio and Modularity metrics improved performance of Conductance considered by itself by 83% on Twitter network ($\rho = 0.413$ vs. $\rho = 0.226$) and by more than 54% on DBLP network ($\rho = 0.327$ vs. $\rho = 0.212$) (cf. Supplementary Table 6 and Supplementary Table 4). This observation suggests that different metrics considered together can more accurately predict community ranks than any one metric by itself.

Second, while performance of baseline community metrics was improved by aggregation, CRank achieved better performance than aggregated baseline community metrics on all five datasets. CRank performed up to 80% better than combinations of baseline metrics and generated on average 38% better community rankings. This result is also interesting because the baselines aggregate five or even six community metrics but CRank aggregates only four CRank metrics (Supplementary Table 6). With these results, we conclude that improvement of CRank's performance does not come solely from the aggregation itself, but rather also from CRank metrics.

Supplementary Table 6: **Performance of CRANK vs. combinations of baseline community metrics.** Each experiment combined baseline community metrics by averaging their scores. We measure Spearman's rank correlation between the generated prioritization of communities and the gold standard ranking of communities. Communities were detected by the CoDA algorithm[9]. Cnd: Conductance[50], Mod: Modularity[51], FOMD: Fraction over median degree[46], TPR: Triangle participation ratio[46], Flake-ODF: Out degree fraction[46]. Higher value indicates better performance.

| Baseline community metrics | Facebook | Amazon | Google+ | Twitter | DBLP | Average |
|---|---|---|---|---|---|---|
| {FOMD, TPR, Cut Ratio, Cnd, Mod} | 0.253 | 0.202 | 0.219 | 0.413 | 0.327 | 0.283 |
| {FOMD, TPR, Cut Ratio, Flake-ODF, Mod} | 0.144 | 0.152 | 0.173 | 0.208 | 0.317 | 0.199 |
| {FOMD, TPR, Flake-ODF, Cnd, Mod} | 0.225 | 0.178 | 0.202 | 0.298 | 0.367 | 0.254 |
| {FOMD, Flake-ODF, Cut Ratio, Cnd, Mod} | 0.228 | 0.211 | 0.211 | 0.408 | 0.379 | 0.287 |
| {Flake-ODF, TPR, Cut Ratio, Cnd, Mod} | 0.204 | 0.207 | 0.226 | 0.351 | 0.333 | 0.264 |
| {FOMD, Flake-ODF, TPR, Cut Ratio, Cnd} | 0.211 | 0.217 | 0.231 | 0.372 | 0.342 | 0.275 |
| {FOMD, Flake-ODF, TPR, Cut Ratio, Cnd, Mod} | 0.249 | 0.213 | 0.233 | 0.364 | 0.381 | 0.288 |
| CRANK | 0.340 | 0.267 | 0.312 | 0.465 | 0.411 | 0.359 |

### Supplementary Note 8.5 Comparison with other rank aggregation approaches

So far, we learned that CRANK metrics are complementary and that each of them contributes to the performance of CRANK. We would also like to understand the role of another component of CRANK, that is, CRANK rank aggregation method.

To assess the contribution of CRANK rank aggregation method to the overall performance of CRANK, we compare CRANK to its simplified version. Simple CRANK considers exactly the same prioritization metrics but aggregates the metrics using a simple quadratic mean. Given a community $C$, simple CRANK computes the aggregated score $R(C)$ for community $C$ as: $R(C) = \sqrt{\sum_f r_f(C)^2/4}$. We observe that CRANK rank aggregation method consistently outperforms quadratic mean by 20-46% on various datasets (Supplementary Table 7).

Next, we test how CRANK rank aggregation method compares against established rank aggregation approaches[53,63]. Recall that rank aggregation is concerned with how to combine several independently constructed rankings into one final ranking that represents the collective opinion of all the rankings[53]. The classical consideration for specifying the final ranking is to maximize the number of pairwise agreements between the final ranking and each input ranking. Unfortunately, this objective, known as the Kemeny consensus, is NP-hard to compute[53,64], which has motivated the development of methods that either use heuristics or aim to approximate the NP-hard objective[52–54,65]. We compare CRANK rank aggregation method with three other rank aggregation methods that offer guarantees on approximating the Kemeny consensus. We consider a 5-approximation algorithm of the Kemeny optimal ranking called Borda's method[52], a 2-approximation Footrule aggregation[53] and a 2-approximation Pick-a-Perm algorithm[54].

Supplementary Table 7: **Performance of CRANK aggregation method and other rank aggregation methods.** Simple CRANK considers exactly the same prioritization metrics as CRANK but aggregates the scores using the quadratic mean. We measure Spearman's rank correlation between the generated prioritization and the gold standard ranking of communities. Communities were detected by the CoDA algorithm[9]. Higher value indicates better performance.

| Rank aggregation | Facebook | Amazon | Google+ | Twitter | DBLP | Average |
|---|---|---|---|---|---|---|
| Borda | 0.291 | 0.242 | 0.234 | 0.417 | 0.387 | 0.314 |
| Footrule | 0.289 | 0.187 | 0.226 | 0.426 | 0.405 | 0.301 |
| Pick-a-Perm | 0.245 | 0.209 | 0.247 | 0.288 | 0.276 | 0.253 |
| Simple CRANK | 0.250 | 0.190 | 0.213 | 0.341 | 0.340 | 0.267 |
| CRANK | 0.340 | 0.267 | 0.312 | 0.465 | 0.411 | 0.359 |

Results in Supplementary Table 7 show that rank aggregation in CRANK is effective as it either matched or outperformed alternative rank aggregation approaches although CRANK does not approximate the Kemeny consensus. CRANK outperformed Borda's method, the best performing alternative approach, by at least 6%. Across all datasets, CRANK achieved 14% higher average Spearman's rank correlation than Borda's method. This observation is interesting, since Borda's method is the most natural and usual choice for rank aggregation[53]. Pick-a-Perm generally performed the worst among the considered methods. Pick-a-Perm operates by returning *one* of the input rankings selected at random. Although it is a 2-approximation algorithm to the Kemeny optimal ranking[52], it may be of limited practical value when the goal is to maximize coherence of the final ranking with *all* the input rankings (which is the case in our study). We note that since finding the optimal Kemeny solution is NP-hard, none of the algorithms, including CRANK, guarantees to provide the optimal solution, and different algorithms typically find different solutions. However, CRANK achieved on average 27% higher Spearman's rank correlation than alternative approaches that combine metric scores by approximating the NP-hard objective.

In addition to consistently producing better results, CRANK rank aggregation method has two important advantages over alternative rank aggregation methods. First, CRANK handles inconsistencies between the ranked lists (*i.e.*, input rankings) by estimating the importance weights for each ranked list. It combines different metrics such that the weight of each metric varies with community rank. As such, CRANK allows a practitioner to explore, for each community, the weight of each metric in the aggregated community ranking. The importance weights also take account of uncertainty in a ranked list. When combining the ranked lists into a final ranking, CRANK uses the weights to down-weight uninformative parts of each ranked list and up-weight informative parts of each ranked list (Supplementary Note 4). Experiments suggest that the importance weight-based approach plays a role in good performance of CRANK.

Second, CRANK rank aggregation method can consider meta or other label information when combining the metrics. This capability is important because meta information can guide the method toward producing more useful results (Supplementary Note 10). This is in sharp contrast with other rank aggregation methods, which are unsupervised methods.

## Supplementary Note 9 Inclusion of additional community metrics into CRANK

So far, we showed that CRANK represents a flexible and general community prioritization platform whose model and metrics capture conceptually distinct network structural features. The metrics non-redundantly quantify different features of network community structure (Supplementary Note 8.3). We also showed that each CRANK metric is necessary and contributes positively to the performance of CRANK (Supplementary Note 8.3). Unlike alternative network metrics, such as conductance, CRANK metrics capture both the magnitude and the robustness of network structural features (Supplementary Note 8.4).

However, it is not possible to theoretically guarantee that any finite set of metrics will be sufficient for prioritizing communities in all real-world networks. We address this challenge by showing how to integrate any number of additional user-defined metrics into CRANK model without requiring further technical changes to the model. This way, CRANK can build on any existing body of network metrics and can consider domain-specific community/cluster metrics.

### Supplementary Note 9.1 Sensitivity of CRANK to adding low-signal community metrics

We performed additional analyses investigating how inclusion of potentially noisy metrics affects CRANK performance.

We created synthetic networks with planted community structure using a stochastic block model. For a given synthetic network we applied a community detection method[9] to detect communities and then used CRANK to prioritize them. We measured prioritization performance using Spearmans rank correlation between CRANK ranking and the gold standard ranking of communities, as described in the manuscript. We repeated the experiment many times, each time adding a different number of noisy metrics to CRANK. Each added metric was a noisy version of the gold standard ranking of communities containing a different amount of useful signal.

We report results in Supplementary Figure 3. We find that CRANK's performance degrades gracefully when low-signal metrics or even adversarial metrics (*i.e.*, metrics that correlate negatively with the gold standard community ranking) are added to the set of metrics aggregated by CRANK (Supplementary Figure 3). For example, adding 6 additional noisy metrics to CRANK, each correlating 0.10 with the gold standard community ranking, improves CRANK performance by 11%.

We also find that CRANK's performance improves substantially when only a relatively few metrics are added to the set of metrics aggregated by CRANK, if the added metrics are positively correlated with the gold standard ranking. For example, adding 3 additional metrics to CRANK, each correlating 0.50 with the gold standard community ranking, improves CRANK's performance by 67% (Spearman's rank correlation $\rho > 0.90$, Supplementary Figure 3).

These analyses show that CRANK can handle a large number of metrics and that its aggregation method is robust to adding low-signal metrics.

## Supplementary Note 10 Integration of domain-specific information into CRANK

Next, we turn our attention to studying how CRANK can incorporate domain-specific (supervised) information in community prioritization. For domains at the frontier of science supervised data is often scarce and thus unsupervised approaches, like CRANK, are extremely important. In domains where domain-specific or other meta and supervised data is available, our method can easily consider such information, potentially leading to improved community prioritization.

In this note, we demonstrate that CRANK has a unique ability to operate in unsupervised as well as supervised environments, and thus can identify high-quality communities when domain-specific information is available and even when it is not.

### Supplementary Note 10.1 Integration of domain-specific information into CRANK

When domain-specific or other meta and label information is available it can prove to be useful to improve prioritization performance. In the context of biological networks, domain-specific information is often given in the form of pathways or gene sets that are over-represented among genes belonging to a cluster/community[66–73]. CRANK can easily use such domain-specific or other meta and label information to supervise community prioritization. When external information about communities is available, CRANK can make advantage of it to boost prioritization performance. CRANK can leverage available meta information at two different stags of analysis as follows.

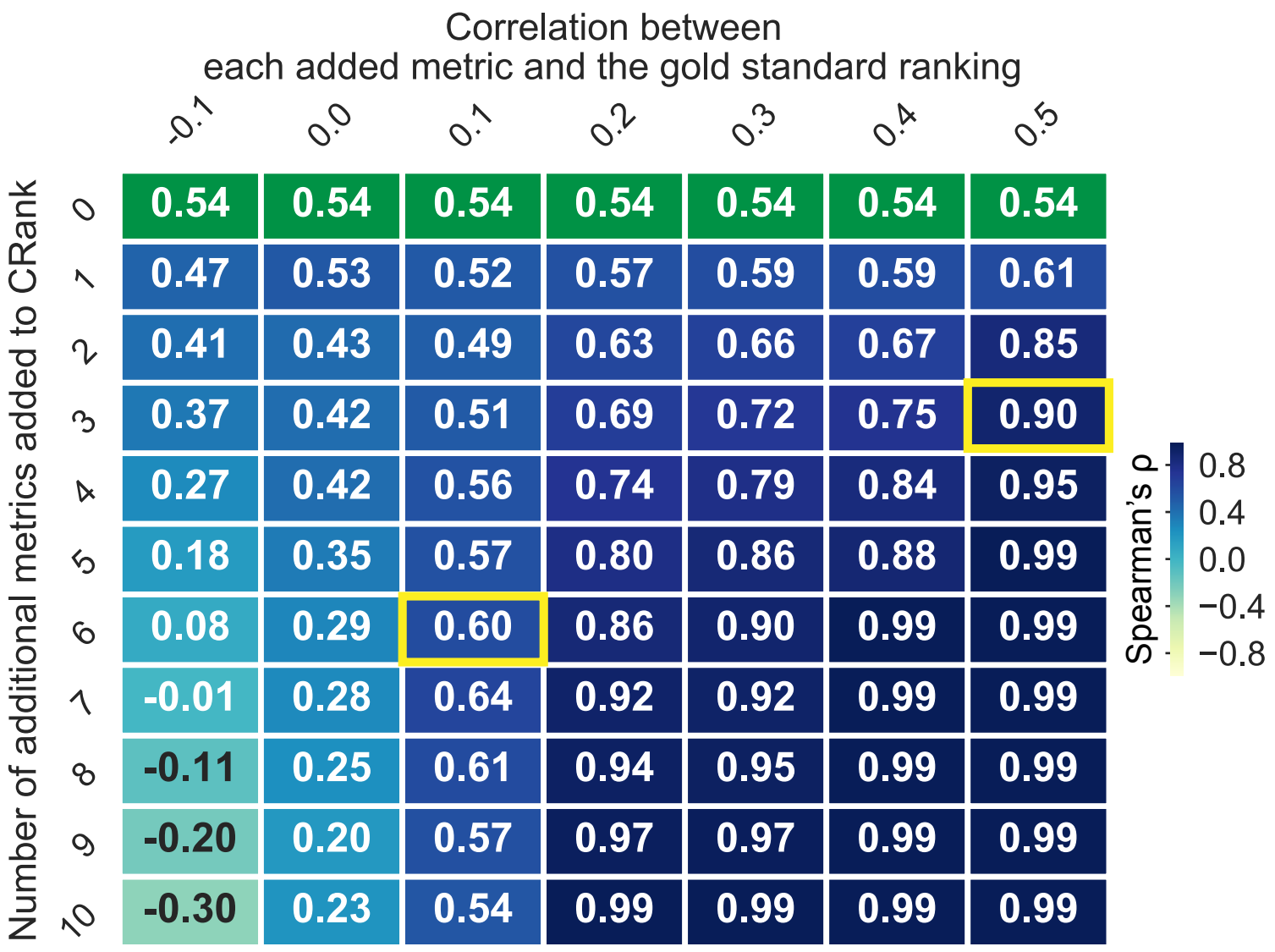

Supplementary Figure 3: **Connections between the number of additional community prioritization metrics in CRank, the signal rate of each added metric, and CRank s performance.** CRank is based on aggregation of four metrics, *i.e.*, Likelihood, Density, Boundary, and Allegiance, that each characterizes a different aspect of community structure. CRank can include additional metrics to the set of aggregated metrics. We create a synthetic network with planted community structure using a stochastic block model with 100 planted modules/communities. We apply a community detection method[9] to detect communities and then use CRank to prioritize them. CRank produces a ranked list of communities. The gold standard rank of each community is determined by how accurately it corresponds to its planted counterpart. We measure prioritization performance using Spearmans rank correlation between CRank ranking and the gold standard ranking of communities. We obtain $\rho = 0.54$ based on CRank metrics (in green). We repeat the experiment many times, each time based on a different number of metrics added to CRank (y-axis in the heat map), where each added metric is a noisy version of the gold standard community ranking with a specific amount of signal (x-axis in the heat map). The heat map shows that adding only a relatively few metrics to the set of metrics aggregated by CRank can lead to an almost perfect prioritization ($\rho > 0.90$) if the added metrics are positively correlated with the gold standard ranking. For example, adding 3 noisy metrics to CRank, each correlating 0.50 with the gold standard community ranking, improves CRank s performance by 67% ($\rho = 0.90$, highlighted cell in the heat map). We observe that CRank's performance degrades gracefully when low-signal metrics or even adversarial metrics (*i.e.*, metrics that correlate negatively with the gold standard community ranking) are added to CRank. For example, adding 6 additional noisy metrics to CRank, each correlating 0.10 with the gold standard community ranking, improves CRank s performance by 11% ($\rho = 0.60$, highlighted cell in the heat map).

**Domain-specific information at network community prioritization stage.** Given side information about a small number of high-quality communities, CRANK can use these high-quality communities to guide the prioritization. We only slightly modify the original algorithm where we use supervised information for CRANK to determine importance weights for each prioritization metric and each bag (Eq. (26) in CRANK algorithm). Importance weights are thus determined in a supervised manner based on the given high-quality communities, such that larger weights are assigned to metrics and bags that contain a larger number of communities with high-quality labels.

**Domain-specific information at network community detection stage.** A complementary approach to integrating meta-information at community prioritization stage is to integrate it at community detection stage. Recent community detection methods[17,19,74] can incorporate metadata into a community detection method itself, which helps guide the method to detect more useful communities. These methods combine network and meta-information about nodes, such as the age of individuals in a social network or mutation effects of genes in a gene network, to improve the quality of detected communities. CRANK can be used in conjunction with those methods.

### Supplementary Note 10.2 Effective use of domain-specific information by CRANK

We have conducted additional analyses on synthetic and real-world networks showing how CRANK can integrate domain-specific information into its prioritization model to boost performance.

**Synthetic networks with planted community structure.** In experiments on synthetic networks with planted community structure, we observe that CRANK can use label information about high-quality communities when calculating importance weights for prioritization metrics. We observe that label information improves CRANK's performance by up to 14–117%, depending on the amount of provided information used for supervision (Supplementary Figure 4).

**Network of medical drugs.** In experiments on the medical drug network, we evaluate CRANK's ability to incorporate information about medical drugs into prioritization of drug communities (Supplementary Figure 5). We find that including drug-specific information significantly improves CRANK's performance, even when the amount of drug-specific information used for supervision is small. Supervised CRANK produces up to 55% better community rankings than can be produced by unsupervised version of CRANK ($\rho = 0.48$ vs. $\rho = 0.31$, left panel; $\rho = 0.47$ vs. $\rho = 0.38$, middle panel; $\rho = 0.61$ vs. $\rho = 0.53$, right panel in Supplementary Figure 5).

These results show that CRANK can identify high-quality communities when meta or other label information is available and even when it is not. Thus, CRANK can operate in supervised and unsupervised environments and effectively prioritize communities. These analyses increase our confidence that CRANK will be of broad practical utility in both domains with abundant and scarce domain-specific knowledge.

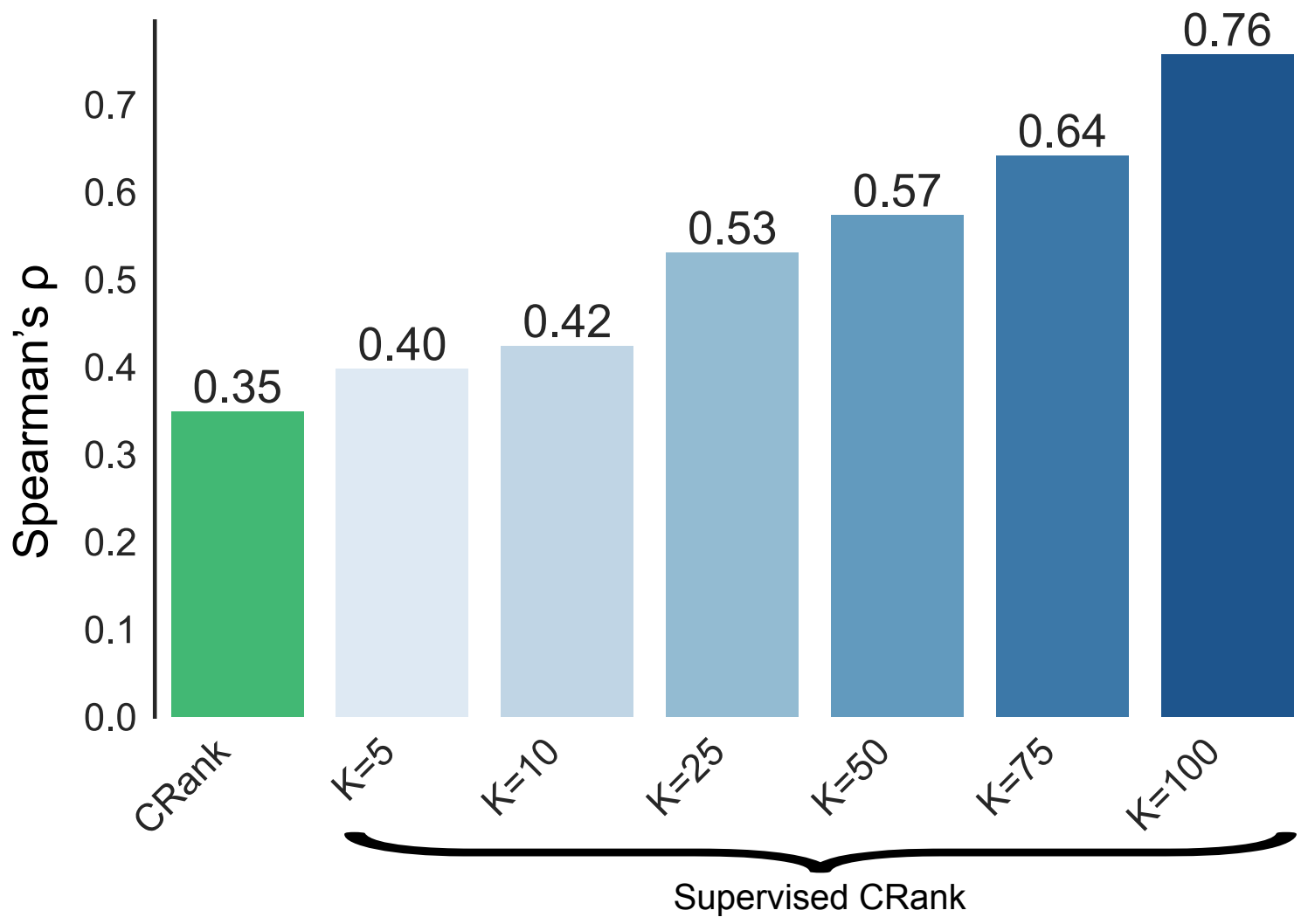

Supplementary Figure 4: **Integrating meta information into CRANK at network community prioritization stage.** CRANK can easily incorporate supervised information or external knowledge as prior or even as supervised labels. On synthetic networks with planted community structure we test how useful supervised information can be for CRANK. We generate a benchmark network on $N = 6000$ nodes using a stochastic block model with 200 planted modules/communities. (The same stochastic block network model is used in experiments reported in Figure 2 in the manuscript.) Each planted community has 30 nodes. Planted communities use different values for within-community edge probability, one hundred use $p = 0.6$ and one hundred use $p = 0.2$. Between-community edge probability is $p = 0.02$. Given a benchmark network, we apply a community detection method to detect communities[9] and then use CRANK to prioritize them. CRANK produces a ranked list of communities that we evaluate against a gold standard community ranking using Spearman's rank correlation, as described in the manuscript. Each bar indicates performance of CRANK in an experiment with a different amount of supervised information. Unsupervised experiment is indicated in green ($\rho = 0.35$). In every other experiment (indicated in increasing shades of blue), CRANK is given supervised information about a set of $K$ high-quality communities and it can use this meta-information for supervised community prioritization. High-quality communities are communities with the highest fraction of nodes correctly classified into their corresponding planted communities. CRANK uses these communities to determine importance weights for each prioritization metric and each bag (Eq. (26) in CRANK algorithm) in a supervised manner, such that larger weights are assigned to metrics and bags that contain a larger number of high-quality communities. Integration of meta-information in CRANK improves its performance by 14-117%, depending on the amount of additional information specified by the size of set $K$.

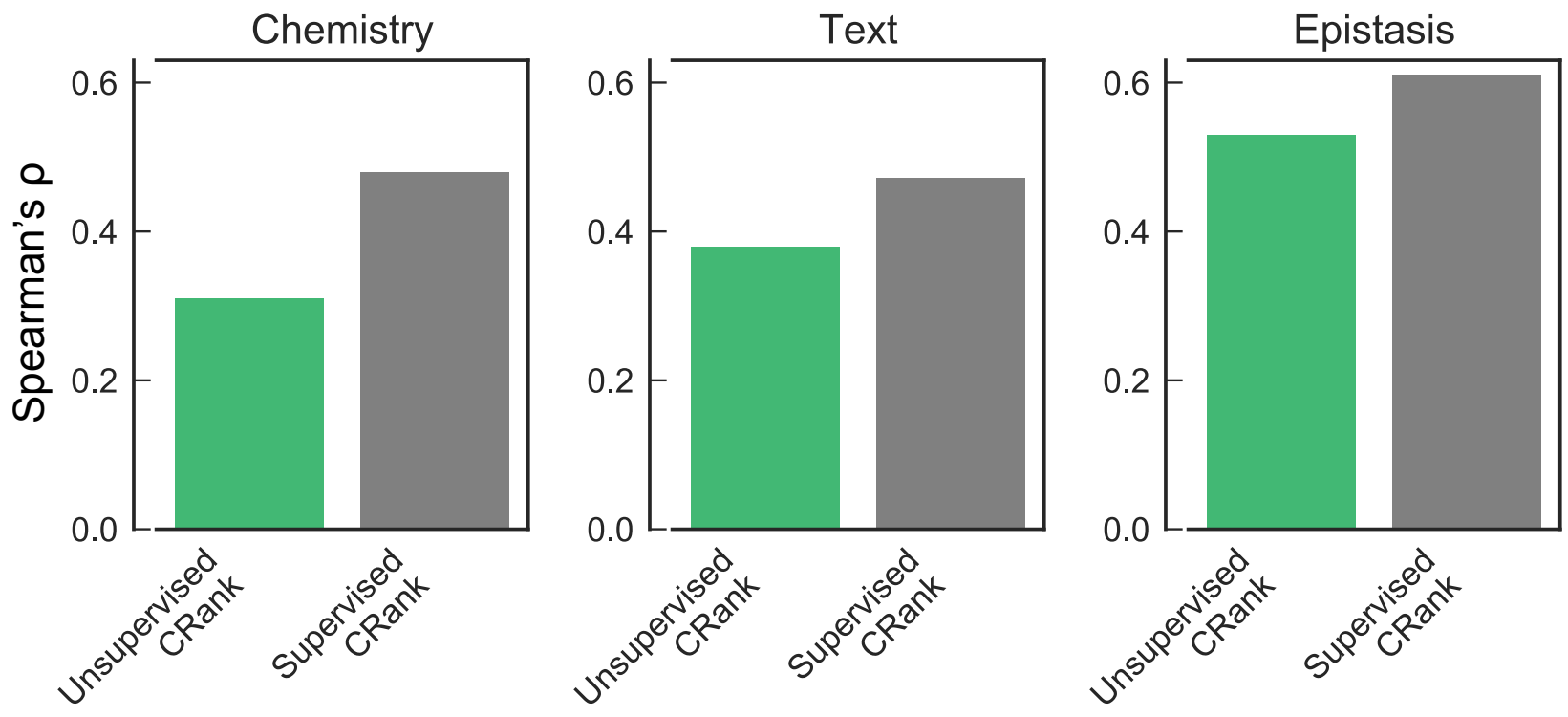

Supplementary Figure 5: **Integration of supervised information about medical drug into CRANK.** We use the network of medical drugs (Figure 3 in the manuscript) to study how useful supervised information about drugs can be for CRANK. Recall that we have three types of external meta-information for each drug community: (1) chemical structure similarity of the drugs ("Chemistry"), (2) associations between drugs derived from text data ("Text"), and (3) drug-drug interactions between the drugs ("Epistasis"). Given the medical drug network, we apply a community detection method to detect communities[9] and then use CRANK to prioritize them. CRANK produces a ranked list of communities that we evaluate against a gold standard community ranking using Spearman's rank correlation, as described in the manuscript. However, in contrast to experiments reported in the manuscript (Figure 3 in the manuscript), we here use the supervised version of CRANK to prioritize communities. Each bar indicates performance of CRANK in an experiment with a different type of supervised information. Unsupervised experiments are indicated in green ($\rho = 0.31$ for Chemistry, $\rho = 0.38$ for Text, and $\rho = 0.53$ for Epistasis, as in Figure 3 in the manuscript). In every other experiment (indicated in grey), CRank is given a set of $K = 10$ high-quality communities determined using an external chemical database (*i.e.*, drug-drug interactions from the Drugbank[36] database). CRANK uses high-quality communities to determine importance weights for each prioritization metric and each bag (Eq. (26) in CRANK algorithm) in a supervised manner, such that larger weights are assigned to metrics and bags that contain a larger number of high-quality communities. Inclusion of drug-specific information significantly improves CRANK's performance, although the amount of supervised information is small. Although in each experiment CRANK has access to only 10 high-quality communities, supervised CRANK produces up to 55% better community rankings than can be produced by unsupervised version of CRANK ($\rho = 0.48$ vs. $\rho = 0.31$, left panel; $\rho = 0.47$ vs. $\rho = 0.38$, middle panel; $\rho = 0.61$ vs. $\rho = 0.53$, right panel).

## Supplementary Note 11 Further case studies

In this note we describe case studies on medical, social, and information networks, beyond those presented in the main text.

## Supplementary Note 11.1 Amazon product co-purchasing network

The CRANK approach also provides new insights into high-quality communities beyond community rankings in biomedical networks. Results on a large network of frequently co-purchased products at the online retailer further underpin the need for automatic community prioritization. We detect communities in the Amazon product network and rank them using CRANK (Supplementary Figure 6). We find that communities ranked high by CRANK mostly contain products that belong to the same product category (Supplementary Figure 6a). For example, the rank 2 community (2nd highest community in the ranking) contains books belonging to a children's literary franchise "The Boxcar Children" about orphaned children who create a home in an abandoned boxcar. Another high-ranked (rank 3) community is about progressive country, a subgenre of country music. In contrast, communities ranked lower by CRANK carry much broader semantic meaning and their products become increasingly more heterogeneous (Supplementary Figure 6a).

## Supplementary Note 11.2 Human symptoms disease network

We consider a symptom-based human disease network[44], where a link between two diseases indicates that they have significantly similar clinical symptoms. Promising disease communities in this network are communities with similar molecular, genetic, and chemical properties because such communities hold promise for development of new therapeutic strategies[75–77]. We apply CRANK to the disease network and examine whether it ranks higher communities that are considered more promising.

The disease network was constructed based on more than seven million PubMed bibliographic records[44]. From these records, the symptom-disease relationships were extracted and the symptom similarities for all disease pairs were quantified resulting in the network with 133,106 connections with positive similarity between 1,596 diseases[44]. The network is visualized in Supplementary Figure 7a. The disease network covers a spectrum of disease categories, from broad categories such as cancer to specific conditions such as hyperhomocysteinemia.

After detecting disease communities using a community detection method[9], we prioritize the communities using CRANK. We then evaluate the degree of correspondence between the CRANK ranking of disease communities and the gold standard ranking. We consider three external medical databases[37] with molecular, genetic, and chemical information about diseases (Supplementary Note 6.2). This way, we obtain three possible gold standard rankings. The gold standard rankings are: (1) the ordering of communities by the overlap in disease-associated molecular pathways, (2) the ordering of communities by the similarity of genes associated with diseases in each commu-nity, and (3) the ordering of communities by the structure similarity of chemicals associated with diseases within each community.

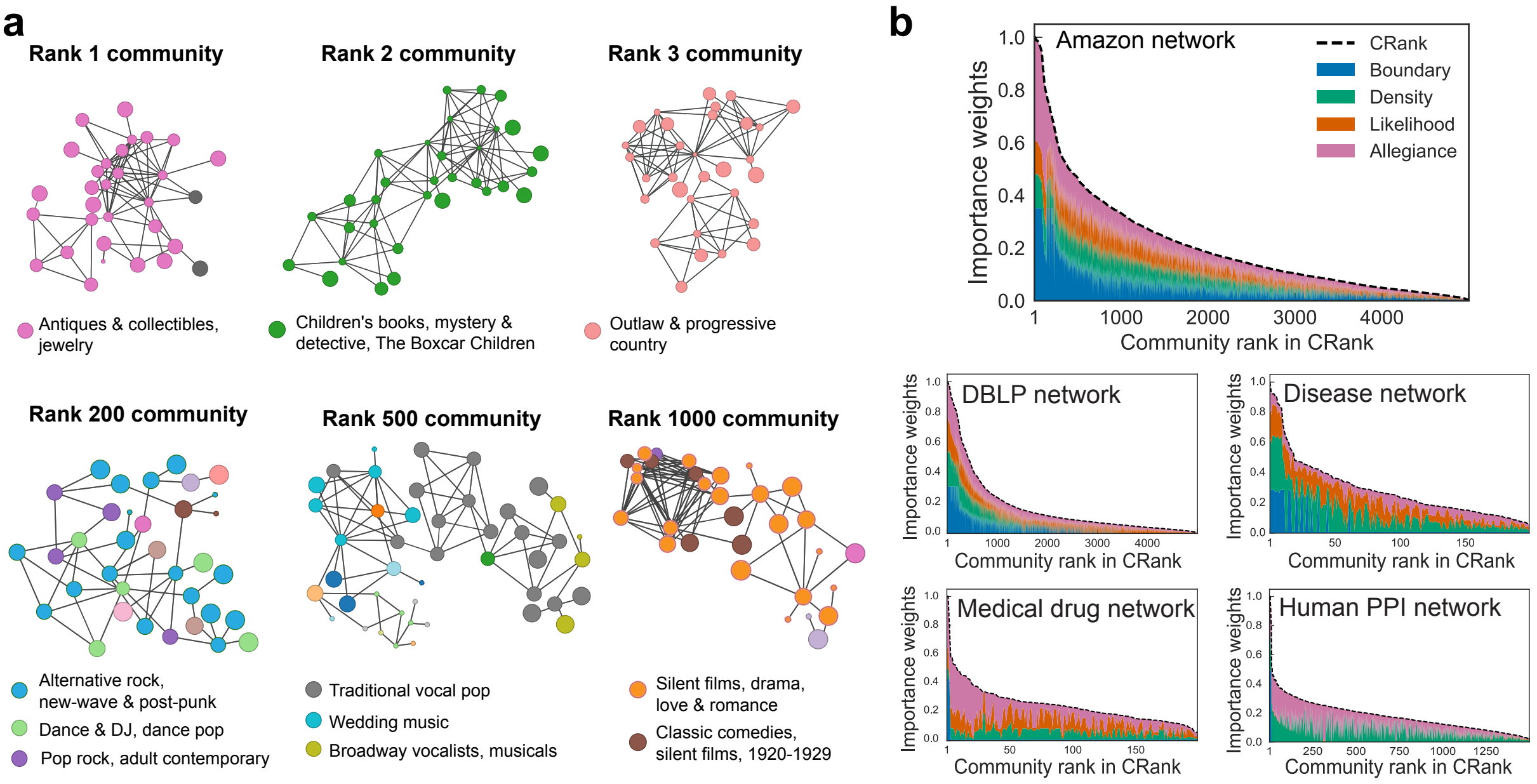

Supplementary Figure 6: **Prioritizing network communities in the Amazon product co-purchasing network and the community prioritization metrics.** **(a)** The network has more than 300,000 nodes and nearly one million edges. Communities are detected using a statistical community detection method[9] (more than 10,000 communities are detected) and prioritized using CRank. Product categories are provided by Amazon and products (nodes) are colored by their category. Manual inspection reveals top ranked communities correspond to the coherent groups of highly related products, such as antique jewelry (rank 1), children's books (rank 2), and country music (rank 3), whereas lower ranked communities contain diverse sets of products from different categories. For example, the rank 200 community contains song albums by bands within the broader rock musical style, such as music by English new wave rock bands and English pop bands, but also pop albums, such as "They Called Him Tintin." The rank 500 community is even less coherent. It contains Broadway musicals, including "House of Flowers" and "Bells Are Ringing," as well as several albums with wedding music, such as "Great Wedding Songs" and "A Song For My Son On His Wedding Day." The rank 1000 community consists predominantly of classic silent films from a wide range of genres, including drama, action, romance, and comedies. **(b)** The importance of prioritization metrics in CRank varies across different networks and across communities within each network. CRank aggregates the values of different metrics such that the weight of a metric varies with community. In the Amazon network, allegiance and boundary metrics are most important for the high-ranked communities, indicating that nodes in the high-ranked communities preferentially attach to other nodes belonging to the same community, and edges connecting each community with the rest of the network are weak. In contrast, density and likelihood metrics contribute less to the prioritization, which means that likelihood and density of a community are less indicative of its quality in the Amazon network.

We evaluate CRANK performance by measuring how well its ranking corresponds to available disease-chemical, disease-gene, and disease-pathway gold standard rankings. We quantify the results using Spearman's rank correlation $\rho$ between the CRANK ranking and the gold standard ranking. The results in Supplementary Figure 7b show that CRANK successfully ordered the communities based on how well they match data in the external medical databases. We observed that CRANK ranking agreed well with the gold standard ranking based on molecular pathways ($\rho = 0.45$, p-value = $1.7 \times 10^{-7}$), genetic associations ($\rho = 0.47$, p-value = $2.7 \times 10^{-8}$), and chemical associations ($\rho = 0.51$, p-value = $2.0 \times 10^{-9}$).

We contrast the ranking provided by CRANK with the ordering of disease communities by Modularity[51]. Modularity-based ranking (Supplementary Figure 7c) achieved Spearman's rank correlation of $\rho = 0.01$ on molecular pathway data, $\rho = 0.16$ on genetic association data, and $\rho = 0.12$ when evaluated against external database with chemical associations. When comparing CRANK with Modularity we see that CRANK ranking is 3- to near 50-fold better than the ranking by Modularity as quantified by Spearman's rank correlation. The result that CRANK's high-ranked communities coincide with groups of diseases with similar genetics is interesting for understanding etiology of diseases, which can help with drug repurposing[77].

An alternative to prioritizing communities based on network structure alone might be to prioritize communities using data in an external medical database. The main obstacle to using external data for community prioritization is that comprehensive and unbiased external data are rarely available in real world. Our analysis of the human disease network involved known diseases for which molecular, genetic or chemical information is available in the medical databases. However, the network of all medical diagnoses contains over one hundred million diagnoses[78] assigned to patients in hospitals, the vast majority of which have yet unknown molecular, genetic or chemical origins. CRANK offers itself as an interesting approach for prioritizing diseases communities in such cases, because CRANK uses only information provided by the network structure.

### Supplementary Note 11.3 Further details on prioritizing drug communities

Beyond results described in the main text, we here report prioritization performance of conductance and test how conductance compares to CRANK on the network of medical drugs. Recall that the network of medical drugs connects two drugs if they share at least one target protein. Supplementary Figure 8 shows that CRANK ranking of drug communities outperforms ranking by conductance on all three types of ground-truth information about chemicals.

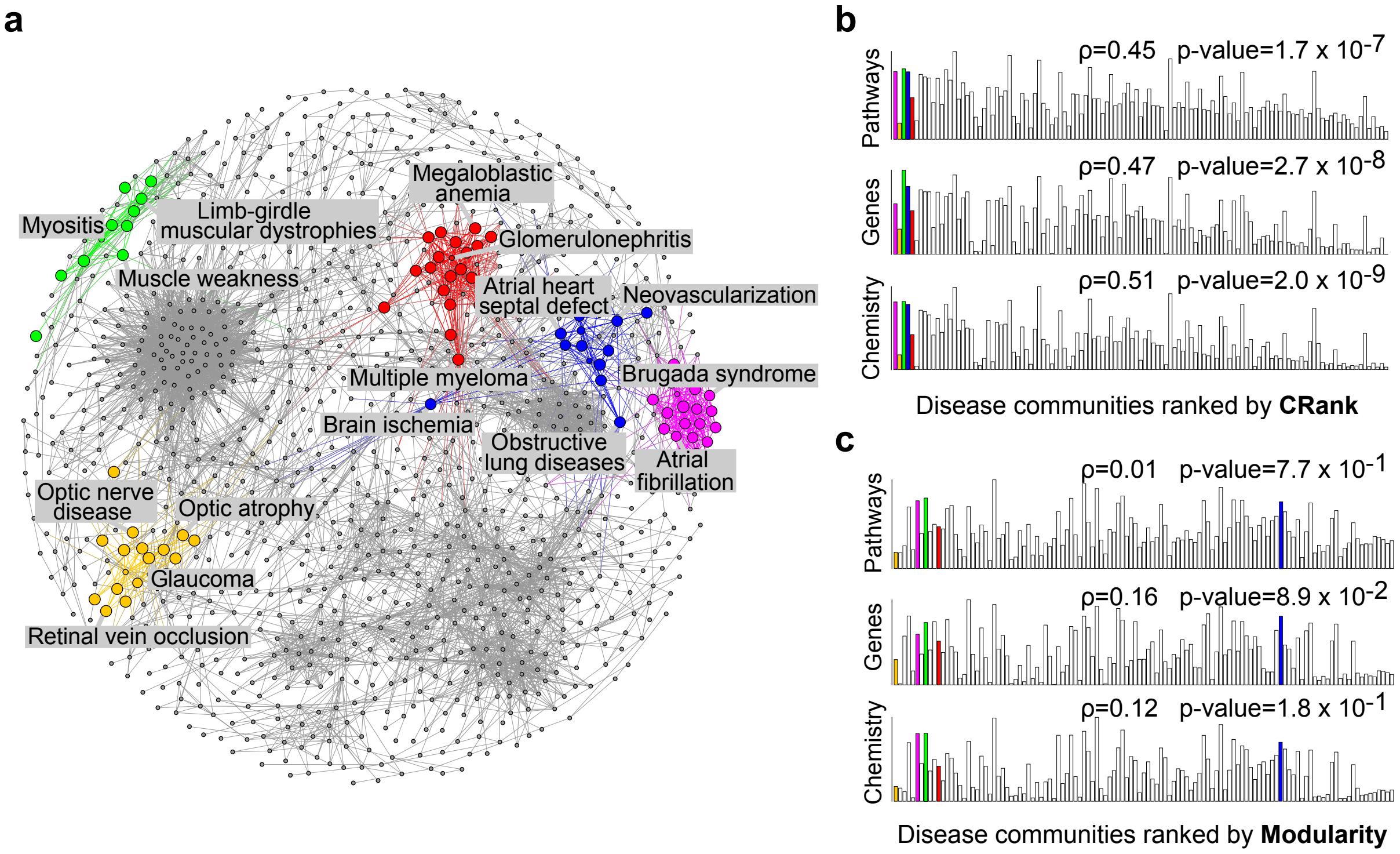

Supplementary Figure 7: **Prioritizing network communities found in the human symptoms disease network.** **(a)** Human symptoms disease network[44] links diseases that have significantly similar clinical manifestation. Highlighted are top five disease communities as determined by CRANK. Nodes of the highlighted communities are sized by their likelihood score (Eq. (5)). **(b-c)** We evaluate community prioritization against three external medical databases that were not used during community detection or prioritization: disease-pathway associations[37] ("Pathways"), disease-gene associations[37] ("Gene"), and disease-chemical associations[37] ("Chemistry"). Bars in barplots represent disease communities; bar height denotes similarity of diseases in a community with regard to an external medical database. In perfect prioritization, the heights of the bars would be decreasing from left to right. **(b)** CRANK ranking exhibits 3- to near 50-fold better correspondence (quantified by Spearman's rank correlation coefficient $\rho = 0.45, 0.47, 0.51$) with the three external medical databases than **(c)** ranking of communities by Modularity ($\rho = 0.01, 0.16, 0.12$).

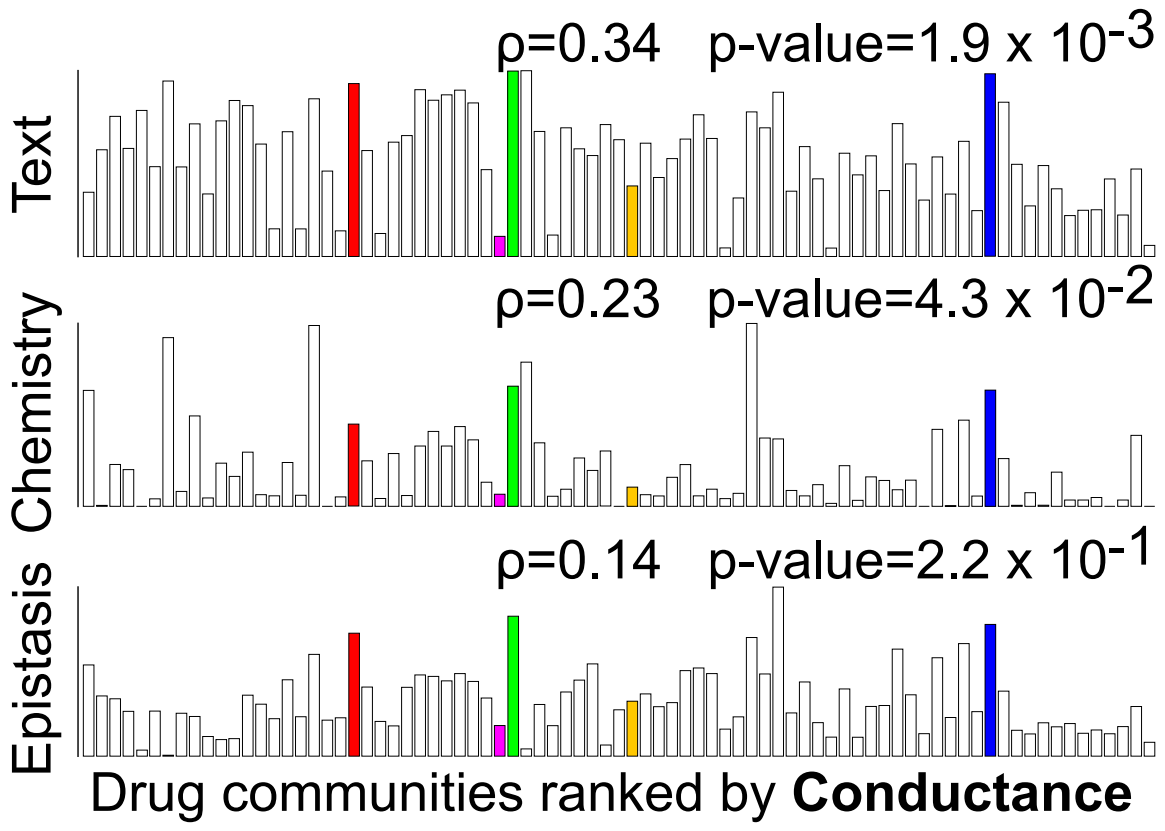

Supplementary Figure 8: **Performance of conductance in the network of medical drugs.** The network of medical drugs connects two drugs if they share at least one target protein. Communities were detected by a community detection method[9], and then prioritized by CRANK or an alternative method, such as conductance. We evaluate community prioritization against three external chemical databases (Supplementary Note 6) that were not used during community detection or prioritization. For each community we measure: (1) drug-drug interactions between the drugs ("Epistasis"), (2) chemical structure similarity of the drugs ("Chemistry"), and (3) associations between drugs derived from text data ("Text"). We expect that a true high-priority community will have more drug-drug interactions, higher similarity of chemical structure, and stronger textual associations between the drugs it contains. Bars represent communities; bar height denotes similarity of drugs in a community with regard to the gold standard based on external chemical databases. In a perfect prioritization, bars would be ordered such that the heights would decrease from left to right. CRANK ranking of drug communities outperforms ranking by conductance across all three chemical databases (as measured by Spearman's rank correlation $\rho$ with the gold standard ranking). CRANK ranking achieves $\rho = 0.38, 0.31, 0.53$ (see main text), while conductance obtains $\rho = 0.34, 0.23, 0.14$.

### Supplementary Note 11.4 Further details on prioritizing gene communities

We apply CRANK to five molecular biology networks describing physical, genetic, and regulatory interactions between genes and proteins (Supplementary Note 6.1). Community detection in such networks is useful because the detected communities tend to correlate with cellular functions, protein complexes and disease pathways[41,79,80], and thus they provide a large pool of candidates out of which relevant communities need to be identified for further biological experimentation.

CRANK takes each network and communities detected in that network[9], and generates a rank-ordered list of communities. Since CRANK ranks the communities purely based on robust-ness and strength of network connectivity, we use the external metadata information about molecular functions, cellular components, and biological processes to assess the quality of community ranking. To this end, we apply statistical enrichment analysis, an established tool in computational biology, to quantify the functional enrichment of each community in molecular functions, components, and processes as captured in the Gene Ontology database[47] (Supplementary Note 7.4). Given a community, the enrichment analysis determines which, if any, of the Gene Ontology terms annotating the genes of the community are statistically over-represented.

Supplementary Table 8: **Prioritization performance in molecular networks.** The fraction of communities that rank among the top 5% and are statistically enriched in molecular functions, biological processes, and cellular components in the Gene Ontology[47] (Supplementary Note 7.4). Higher values are better because they indicate that a higher fraction of top-ranked communities achieve significant correspondence with external knowledge in the Gene Ontology. Communities were detected using a statistical community detection method[9] and prioritized using one of four different approaches.

| Method | Human Net | Human IntAct | Yeast GI | Human BioGRID | Human STRING |
|---|---|---|---|---|---|
| Random | 0.104 | 0.216 | 0.227 | 0.128 | 0.125 |
| Modularity | 0.632 | 0.587 | 0.598 | 0.597 | 0.624 |
| Conductance | 0.658 | 0.644 | 0.688 | 0.523 | 0.518 |
| CRANK | 0.811 | 0.707 | 0.747 | 0.689 | 0.691 |

We measure if the highest ranked communities in each network are more enriched in the GO terms than what would be expected by chance. Supplementary Table 8 shows how many communities that rank among the top 5% of all communities in each network are functionally enriched. CRANK ranking contains on average 5 times more communities significantly enriched for cellular functions, components, and processes than random prioritization, and 13% more significantly enriched communities than modularity or conductance-based ranking.

For example, a community detection method[9] detected 1,500 communities in the human protein-protein interaction (PPI) network. CRANK prioritized the communities by producing a rank-ordered list of all detected communities in the network. Supplementary Table 9 shows ten highest ranked communities by CRANK. The highest ranked community is composed of 20 genes, including *PORCN*, *AQP5*, *FZD6*, *WNT1*, *WNT2*, *WNT3*, and other members of the Wnt signaling protein family[81]. Genes in that community are enriched in the Wnt signaling pathway processes (p-value = $6.4 \times 10^{-23}$), neuron differentiation (p-value = $1.6 \times 10^{-15}$), cellular response to retinoic acid (p-value = $2.9 \times 10^{-14}$), and in developmental processes (p-value = $9.2 \times 10^{-10}$), among others.

These results highlight the potential of CRANK to aid in the identification of relevant communities from a large pool of communities detected in molecular networks.

Supplementary Table 9: **Highest ranked gene communities in the human PPI network.** A human PPI network was compiled using interaction data from the STRING database[43]. A statistical community detection method[9] was used to detect communities in the network, followed by CRANK to prioritize the communities. Listed are ten highest ranked communities. Functional enrichment of each community was determined by performing gene set enrichment analysis (Supplementary Note 7.4) based on the Gene Ontology term-associated gene sets[47].

| CRANK | Community | Statistically over-represented Gene Ontology terms |
|---|---|---|
| Rank 1 | *WNT2B, PORCN, AQP5, FZD6, WNT3, WNT10A, WNT1, WNT6, WNT7A, WNT4, FZD8, WNT2, WNT5B, WNT16, WNT5A, ENSP00000345785, WNT3A, WNT10B, WNT11, WNT7B* | Wnt signaling pathway (p=6.37E-23), Neuron differentiation (p=1.60E-15), Cellular response to retinoic acid (p=2.91E-14), Response to retinoic acid (p=7.73E-13), Canonical Wnt signaling pathway (p=1.84E-12), Cellular response to acid chemical (p=5.43E-11), Single-organism developmental process (p=4.05E-10), Developmental process (p=9.21E-10), Response to acid chemical (p=1.41E-09), Cell surface receptor signaling pathway (p=4.36E-09), Cell differentiation (p=4.54E-09), Anatomical structure morphogenesis (p=9.65E-09), Cellular developmental process (p=3.04E-08), Anatomical structure development (p=4.61E-08), Cellular response to lipid (p=7.58E-08), Cellular response to organic substance (p=1.13E-06), Cell proliferation (p=1.80E-06), Regulation of canonical Wnt signaling pathway (p=2.26E-06), Cellular response to transforming growth factor beta stimulus (p=7.70E-06), Cellular response to stimulus (p=8.87E-06), Response to transforming growth factor beta (p=1.25E-05), Cellular response to chemical stimulus (p=1.34E-05), Mammary gland epithelium development (p=1.46E-05), Regulation of Wnt signaling pathway (p=1.65E-05), Response to lipid (p=1.79E-05), Signal transduction (p=3.82E-05), Response to organic substance (p=1.37E-04), Cellular response to oxygen-containing compound (p=1.73E-04), Chondrocyte differentiation (p=2.03E-04), Epithelium development (p=7.54E-04), Lens fiber cell development (p=8.16E-04), Positive regulation of dermatome development (p=8.16E-04), Regulation of dermatome development (p=8.16E-04), Response to chemical (p=1.58E-03), Stem cell proliferation (p=2.86E-03), Lens development in camera-type eye (p=3.26E-03), Palate development (p=3.89E-03), Response to stimulus (p=4.94E-03), Positive regulation of canonical Wnt signaling pathway (p=5.17E-03), Response to oxygen-containing compound (p=5.73E-03), Animal organ development (p=7.57E-03), Hematopoietic stem cell proliferation (p=8.12E-03), Cellular response to growth factor stimulus (p=2.27E-02), Tissue development (p=2.75E-02), Positive regulation of Wnt signaling pathway (p=3.51E-02), Positive regulation of signal transduction (p=3.66E-02), Neural precursor cell proliferation (p=4.51E-02), Negative regulation of canonical Wnt signaling pathway (p=4.84E-02), Receptor agonist activity (p=1.02E-04), Receptor activator activity (p=9.51E-04), Receptor regulator activity (p=2.05E-03) |
| Rank 2 | *NMUR2, NMUR1, NMU, ENSP00000409127, HCRTR1, NPFFR2, NTSR1, ENSP00000358511, CD200, HCRT, HCRTR2* | Neuropeptide receptor activity (p=8.48E-03), Neuromedin U receptor activity (p=4.51E-02) |
| Rank 3 | *GHRHR, GHRH, GHRL, CCL19, LEP, CCR9, CCL21, ACE2, GHSR, MLNR, CCL25, CXCL13, ENSP00000266003* | Feeding behavior (p=4.73E-05), Adult feeding behavior (p=7.44E-04), Response to hormone (p=2.81E-03), G-protein coupled receptor signaling pathway (p=3.61E-03), Positive regulation of multicellular organism growth (p=3.71E-03), Regulation of response to food (p=3.71E-03), Positive regulation of developmental growth (p=5.48E-03), Regulation of appetite (p=6.48E-03), Positive regulation of response to external stimulus (p=1.15E-02), Positive regulation of cell adhesion (p=4.53E-02), Behavior (p=4.84E-02), Regulation of developmental growth (p=4.84E-02), Ccr chemokine receptor binding (p=1.32E-05), G-protein coupled receptor binding (p=1.56E-04), Ccr10 chemokine receptor binding (p=1.86E-04), Chemokine activity (p=5.08E-04), Chemokine receptor binding (p=1.13E-03), Cytokine activity (p=8.66E-03), Cytokine receptor binding (p=1.48E-02), Receptor binding (p=2.36E-02) |
| Rank 4 | *HLA-DPB1, TRH, ROBO2, CPE, CA12, TRHR, POLD4, RDH11, SLIT1, SLIT3, SLIT2* | Apoptotic process involved in luteolysis (p=1.85E-04), Axon guidance (p=3.70E-04), Neuron projection guidance (p=6.71E-04), Apoptotic process involved in development (p=1.85E-03), Neuron projection extension involved in neuron projection guidance (p=6.45E-03), Axon extension involved in axon guidance (p=6.45E-03), Axon extension (p=1.03E-02), Negative chemotaxis (p=1.03E-02), Neuron projection extension (p=1.54E-02), Developmental cell growth (p=3.02E-02), Ovulation cycle process (p=4.02E-02) |
| Rank 5 | *ENSP00000380280, ENSP00000264498, STAT3, FGF17, ENSP00000260795, KL, FGF19, FGF18, FGF9, FGF8, FGF7, FGF6, FGF5, FGF4, FGF3, FGFR4, FGF1* | Fibroblast growth factor receptor signaling pathway (p=1.21E-19), Transmembrane receptor protein tyrosine kinase signaling pathway (p=1.84E-13), Enzyme linked receptor protein signaling pathway (p=1.49E-11), Positive regulation of cell proliferation (p=2.63E-07), Cell surface receptor signaling pathway (p=5.05E-07), Regulation of cell proliferation (p=2.87E-05), Signal transduction (p=6.69E-04), Regulation of endothelial cell chemotaxis to fibroblast growth factor (p=1.23E-03), Regulation of cell chemotaxis to fibroblast growth factor (p=1.23E-03), Positive regulation of biological process (p=1.10E-02), Regulation of steroid biosynthetic process (p=3.65E-02), Regulation of endothelial cell chemotaxis (p=3.65E-02), Fibroblast growth factor receptor binding (p=1.03E-08), Growth factor receptor binding (p=1.05E-05), Type 1 fibroblast growth factor receptor binding (p=3.08E-04), Type 2 fibroblast growth factor receptor binding (p=3.08E-04) |
| Rank 6 | *KCNH1, ENSP00000222812, KCNG1, KCNG2, KCNG3, KCNG4, KCNV1, KCNS3, KCNB1, KCNV2, ENSP00000254976* | Cellular potassium ion transport (p=1.05E-05), Potassium ion transmembrane transport (p=1.05E-05), Potassium ion transport (p=1.52E-05), Monovalent inorganic cation transport (p=3.99E-04), Inorganic cation transmembrane transport (p=1.66E-03), Cation transmembrane transport (p=1.66E-03), Inorganic ion transmembrane transport (p=3.15E-03), Ion transmembrane transport (p=4.23E-03), Transmembrane transport (p=1.12E-02), Metal ion transport (p=1.17E-02), Cation transport (p=2.61E-02), Delayed rectifier potassium channel activity (p=2.72E-08), Voltage-gated potassium channel activity (p=1.66E-06), Potassium channel activity (p=1.05E-05), Voltage-gated cation channel activity (p=2.14E-05), Potassium ion transmembrane transporter activity (p=2.52E-05), Voltage-gated channel activity (p=6.09E-05), Voltage-gated ion channel activity (p=6.09E-05), Monovalent inorganic cation transmembrane transporter activity (p=5.21E-04), Cation channel activity (p=1.08E-03), Gated channel activity (p=1.55E-03), Metal ion transmembrane transporter activity (p=3.55E-03), Ion channel activity (p=5.27E-03), Inorganic cation transmembrane transporter activity (p=5.57E-03), Substrate-specific channel activity (p=7.23E-03), Channel activity (p=8.82E-03), Passive transmembrane transporter activity (p=8.82E-03), Cation transmembrane transporter activity (p=1.23E-02) |
| Rank 7 | *ENSP00000347979, SPTBN4, KCNH1, KCNC1, KCNG3, ENSP00000345751, ANK3, TIAM1, SCN2A, CNTN1, SCN1B, FADD, ENSP00000376966, SCN5A, KCNV2* | Monovalent inorganic cation transport (p=6.90E-07), Inorganic cation transmembrane transport (p=5.24E-06), Cation transmembrane transport (p=5.24E-06), Inorganic ion transmembrane transport (p=1.30E-05), Ion transmembrane transport (p=1.97E-05), Transmembrane transport (p=7.80E-05), Metal ion transport (p=8.32E-05), Cation transport (p=2.58E-04), Regulation of sodium ion transport (p=3.18E-03), Sodium ion transmembrane transport (p=3.18E-03), Ion transport (p=4.97E-03), Sodium ion transport (p=5.62E-03), Establishment of localization (p=3.12E-02), Cation channel complex (p=2.59E-06), Sodium channel complex (p=8.17E-06), Ion channel complex (p=2.03E-05), Node of ranvier (p=3.80E-05), Transmembrane transporter complex (p=6.77E-05), Transporter complex (p=7.54E-05), Cell-cell contact zone (p=1.27E-03), Axon part (p=2.58E-03), Plasma membrane part (p=3.02E-03), T-tubule (p=2.01E-02), Membrane protein complex (p=2.62E-02), Voltage-gated channel activity (p=6.24E-06), Voltage-gated ion channel activity (p=6.24E-06), Monovalent inorganic cation transmembrane transporter activity (p=8.38E-05), Cation channel activity (p=2.01E-04), Gated channel activity (p=3.11E-04), Metal ion transmembrane transporter activity (p=8.43E-04), Ion channel activity (p=1.35E-03), Inorganic cation transmembrane transporter activity (p=1.44E-03), Substrate-specific channel activity (p=1.97E-03), Channel activity (p=2.50E-03), Passive transmembrane transporter activity (p=2.50E-03), Cation transmembrane transporter activity (p=3.71E-03), Voltage-gated sodium channel activity (p=2.01E-02), Voltage-gated ion channel activity involved in regulation of postsynaptic membrane potential (p=2.01E-02), Ion transmembrane transporter activity (p=2.74E-02), Delayed rectifier potassium channel activity (p=2.87E-02), Sodium channel activity (p=3.94E-02), Substrate-specific transmembrane transporter activity (p=4.89E-02) |
| Rank 8 | *AP2M1, FBXW11, STAT2, CRKL, ENSP00000329418, IFNA2, IFNA1, IFNB1, IFNAR2, ENSP00000337825, IFNAR1, PTPRC, ZAP70, JAK1, TYK2, IFNA8, IRF9, JAK3* | Negative regulation of adaptive immune response based on somatic recombination of immune receptors built from immunoglobulin superfamily domains (p=2.26E-04), Negative regulation of adaptive immune response (p=2.26E-04), Cytokine-mediated signaling pathway (p=2.36E-04), Type I interferon signaling pathway (p=5.76E-04), Negative regulation of leukocyte mediated immunity (p=1.27E-03), Negative regulation of T cell mediated immunity (p=5.73E-03), Regulation of T cell differentiation (p=1.84E-02), Regulation of lymphocyte differentiation (p=3.02E-02), Negative regulation of immune response (p=4.69E-02), Negative regulation of immune effector process (p=4.69E-02), Negative regulation of lymphocyte differentiation (p=4.77E-02), Negative regulation of T cell differentiation (p=4.77E-02) |
| Rank 9 | *ENSP00000372815, CR1, C3AR1, ENSP00000396688, CST3, MASP2, C5AR2, C4B, KDM6B, C3, HCK, APOA2* | Complement activation (p=2.01E-03), Protein activation cascade (p=1.20E-02), Complement binding (p=2.01E-03) |
| Rank 10 | *HSPB1P1, G6PD, TALDO1, MRE11, SLC25A1, HUWE1, MPI, ENSP00000344818, TKT, RPE, DTYMK, KYAT1, KYAT3, NPPA* | Glyceraldehyde-3-phosphate metabolic process (p=1.39E-03) |

## Supplementary references